\documentclass[preprint,12pt]{elsarticle}

\usepackage{amssymb}

\usepackage[utf8]{inputenc} 
\usepackage[T1]{fontenc}    
\usepackage{times}        
\usepackage{hyperref}       
\usepackage{url}            
\usepackage{booktabs}       
\usepackage{amsfonts}       
\usepackage{amsmath}        
\usepackage{amssymb}
\usepackage{amsthm}
\usepackage{mathtools}
\usepackage{bm}             
\usepackage{siunitx}
\DeclareSIUnit{\volpercent}{vol.\%}
\usepackage{nicefrac}       
\usepackage{microtype}      
\usepackage{lipsum}		
\usepackage{graphicx}
\usepackage[]{placeins} 
\usepackage{tabularx}
\usepackage{multirow}

\def\AA{\bm{A}}
\def\bb{\bm{b}}
\def\CC{\bm{C}}

\def\FF{\bm{F}}

\def\ff{\bm{f}}

\def\llt{\bm{l}}

\def\MM{\bm{M}}

\def\rr{\bm{r}}

\def\SS{\bm{S}}

\def\vv{\bm{v}}

\def\unittensor{\bm{i}}

\def\sigmaa{\bm{\sigma}}

\def\tr{\mathrm{tr}}

\def\dev{\mathrm{dev}}

\def\tauu{\mathbb{\tau}}
\def\LLinv{\mathcal{L}^{-1}}
\newcommand{\bkt}[1]{\left(#1\right)}
\newcommand{\Bkt}[1]{\left[#1\right]}

\def\FFT{\bm{F}^\mathrm{T}}

\def\tr{\mathrm{tr}}

\def\dV{\,\mathrm{d}V}
\def\dev{\mathrm{dev}}

\def\tauu{\bm{\tau}}
\def\FFl{\bm{F}^\mathrm{l}}
\def\FFr{\bm{F}^\mathrm{r}}
\def\FFe{\bm{F}^\mathrm{e}}

\def\FFp{\bm{F}^\mathrm{p}}

\def\Psie{\Psi^{\mathrm{e}}}

\def\FFT{\FF^{\mathrm{T}}}

\def\CCe{\CC^\mathrm{e}}
\def\CCl{\CC^\mathrm{l}}

\def\DD{\bm{D}}
\def\DDp{\DD^\mathrm{p}}

\def\DDr{\DD^\mathrm{r}}

\def\mub{\mu^{\mathrm{b}}}
\def\mup{\mu^{\mathrm{p}}}

\def\Ne{N_{\mathrm{e}}}
\def\dT{\dot{T}_{\mathrm{c}}}

\def\Tg{T_{\mathrm{g}}}

\usepackage{xcolor}
\usepackage[
	margin=10pt,
	format=plain,
	font=small,
	labelfont=bf,
	labelsep=colon,
	skip=0pt,
]{caption}

\usepackage[super]{nth}
\usepackage{enumitem}

\usepackage[font=normalsize]{subcaption}
\usepackage{chemmacros}

\newdimen\imageheight

\journal{}

\begin{document}

\begin{frontmatter}



\title{Rate- and temperature-dependent strain hardening in glassy polymers: Micromechanisms and constitutive modeling based on molecular dynamics simulations}


\author{Wuyang Zhao}
\ead{wuyang.zhao@fau.de}

%

\address{Institute of Applied Mechanics, Friedrich-Alexander-Universit\"at Erlangen-N\"urnberg, Egerlandstra{\ss}e 5, 91058 Erlangen, Germany}


\begin{abstract}
We perform molecular dynamics simulations under uniaxial tension to investigate the micromechanisms underlying strain hardening in glassy polymers. By decomposing the stress into virial components associated with pair, bond, and angle interactions, we identify the primary contributors to strain hardening as the stretching of polymer bonds. Interestingly, rather than the average bond stretch, we find that the key contributions to stress response come from a subset of bonds at the upper tail of the stretch distribution. Our results demonstrate that the stress in the hardening region can be correlated with the average stretch of the most extended bonds in each polymer chain, independent of temperatures and strain rates. These bonds, which we denote as load-bearing bonds, allow us to define a local load-bearing deformation gradient in continuum mechanics that captures their contribution to the hardening stress tensor. Building on this insight, we incorporate the load-bearing mechanism into a constitutive framework with orientation-induced back stress, developing a model that accurately reproduces the stress response of the molecular systems over a wide range of temperatures and strain rates in their glassy state.
\end{abstract}




\end{frontmatter}


\section{Introduction}\label{sec:introduction}
Glassy polymers are essentially amorphous polymers at temperatures below their glass transition temperature $\Tg$, exhibiting solid-like mechanical behavior due to restricted atomic mobility from neighboring interactions. The polymer molecules, typically with chain-like structures, can undergo strain hardening at large deformations, effectively increasing the toughness of the materials. Accurately modeling the mechanical response in the hardening region is essential for studying the failure mechanisms of glassy polymers. Unlike rubbers, which are amorphous polymers above $\Tg$, the strain hardening behavior of glassy polymers is sensitive to strain rate and temperature as evident in various experiments \cite{Wendlandt2005,Senden2010,Senden2012} and molecular dynamics (MD) simulations \cite{Hoy2006,Hossain2010}. However, the mechanisms underlying the rate- and temperature-dependence of strain hardening remain not fully understood.

For decades, the hardening mechanism in glassy polymers has often been interpreted through entropic elasticity theory developed for rubbers \cite{Treloar2005,Haward2012}, which attributes stress increases to the reduction of entropy during the elongation of polymer chains. While this theory applies well to rubbers, various experiments \cite{vanMelick2003a,Tian2018} 
and MD simulations \cite{Hoy2006} reveal inconsistencies when applied to glassy polymers. This discrepancy probably arises from a core assumption in entropic elasticity theory that polymer segments are free to adjust their conformations to accommodate deformation. However, in glassy polymers, atomic mobility is restricted, preventing such free adjustment. Consequently, the relaxation time associated with the evolution of polymer segments becomes non-negligible, leading to rate-dependent behavior in the hardening region of glassy polymers. This limited mobility also introduces temperature sensitivity, as relaxation times are typically temperature-dependent. Thus, understanding strain hardening in glassy polymers involves addressing two key questions: (i) which structures primarily contribute to the stress response, and (ii) the relaxation of which structures most significantly affects stress behavior?

To address the first question, various assumptions have been tested via constitutive modeling, given that experimental observation of the microscopic structure during deformation is challenging. Although the precise origin of hardening stress remains unclear, most constitutive models adopt the Lee-Kröner decomposition $\FF=\FFe\FFp$ \cite{Kroener1959,Lee1969} of the deformation gradient $\FF$, where $\FFe$ and $\FFp$ represent the elastic and viscoplastic components of $\FF$, respectively. In this framework, $\FFe$ is assumed to contribute to the stress, which drives the evolution of $\FFp$, characterized by the viscosity $\eta$. To capture strain hardening, Arruda et al. \cite{Arruda1993} introduced a back stress term to account for chain orientation effects. Anand et al. \cite{Anand2003,Anand2009} generalized this formulation by incorporating a unimodular orientation tensor $\AA$ as an internal variable of the free energy density for orientation and proposed that the evolution of $\AA$ is subjected to a dynamic recovery term driven by the back stress. Additional approaches have also been proposed, including models with deformation-dependent viscosity \cite{Wendlandt2005,Senden2012}, a viscous model that uses strain rate as an external variable to capture steady-state mechanical behavior \cite{Zhao2024a}, and models that represent orientation through shear transformation zones (STZs) \cite{Voyiadjis2016,Zhu2022,Lin2022}. While these models successfully replicate the stress response of glassy polymers in the hardening region under various conditions, they remain largely phenomenological, each grounded in different assumptions about the origin of hardening stress. A deeper understanding of the underlying micromechanisms is needed to further elucidate the nature of strain hardening in glassy polymers.

In addressing the second question, it is critical to determine whether the primary relaxation mechanism influencing strain hardening occurs at the scale of polymer chains, such as disentanglement, or at the scale of local segments, like bond rotations. MD simulations have been used to study the mechanisms of disentanglement by tracking the evolution of the entanglement lengths $\Ne$ during deformation \cite{Hoy2006,Jatin2014,Zhu2022}. Hoy and Robbins \cite{Hoy2006} observed that hardening increases with decreasing $\Ne$ only when $\Ne$ is relatively small; beyond a critical value $N^{\mathrm{cr}}_{\mathrm{e}}$, hardening saturates with $\Ne$. This suggests that disentanglement is relevant primarily at smaller $\Ne$. This finding may help to reconcile the conflicting results reported in \cite{Jatin2014} and \cite{Zhu2022}, where $\Ne$ remains constant in \cite{Jatin2014} but evolves in \cite{Zhu2022} during deformation. Furthermore, the critical value $N^{\mathrm{cr}}_{\mathrm{e}}$ is likely temperature-dependent, as increased temperatures accelerate polymer chain relaxation. This implies that disentanglement only occurs at sufficiently high temperatures for glassy polymers, while at low temperatures, the dominant relaxation mode is likely limited to local segmental rotations. These include backbone chain rotations ($\alpha$-relaxation) or side group rotations ($\beta$-relaxation) for certain glassy polymers \cite{BauwensCrowet1969,BauwensCrowet1973,Siviour2005}. Experimental studies on the time-temperature superposition of glassy polymers at large strains \cite{Diani2015,Federico2018,Bernard2020} support this assumption. These studies demonstrate that the stress-strain curves of glassy polymers at specific temperatures and strain rates can coincide in the hardening region, with derived shift factors matching those from small strains. This suggests that similar relaxation mechanisms are active across both small and large strain regimes in these materials. Since disentanglement is not the primary relaxation mechanism at small strains before yielding, it is likely not the dominant mechanism at large strains in the hardening region either, based on these experimental results. Using identical shift factors for both small and large strains, Xiao and Tian \cite{Xiao2019} employed the back stress model \cite{Anand2009} to successfully predict pre-deformation effects in the strain hardening region of glassy polymers.

This paper aims to investigate the microscopic mechanisms underlying strain hardening in glassy polymers using MD simulations, with a focus on capturing its rate- and temperature-dependent behavior. To achieve this, we first decompose the virial stress components into contributions from pair, bond, and angle interactions, identifying bond and angle stresses as the primary contributors to strain hardening. Subsequently, we examine the microscopic quantities that can be correlated with these bond and angle stresses within the hardening region. Our key finding is that these stress terms are correlated with the averaged values of the largest bond stretch in each polymer chain, independent of temperature and strain rate. We denote these bonds as load-bearing bonds and represent this microscopic quantity through a deformation gradient, capturing the macroscopic local load-bearing deformation, denoted as $\FFl$. Building on this result, we propose a decomposition of the total deformation gradient, $\FF=\FFl\FFr$, where $\FFr$ represents the resisting component of the deformation gradient. Finally, we develop a constitutive model within the framework established by \cite{Anand2009}, incorporating orientation-induced back stress. Compared to the model proposed by \cite{Anand2009} and various other hardening models \cite{Xiao2019,Wendlandt2005,Senden2012,Voyiadjis2016,Zhu2022,Lin2022,Zhao2024a}, the improvement presented in this paper lies in providing a clear physical basis for the stress response in the hardening region, as well as a well-defined interpretation of the deformation gradient decomposition $\FF=\FFl\FFr$, where the contribution of $\FFl$ to the stress can be directly extracted from MD simulations.

The remaining part of this paper is organized as follows. In Section \ref{sec:method}, we introduce the MD systems used in this study. Section \ref{sec:MD} presents the analysis of the microscopic mechanisms underlying strain hardening, followed by a discussion on the decomposition $\FF=\FFl\FFr$ in Section \ref{sec:decomposition} based on the findings from this analysis. Section \ref{sec:constitutive} derives the constitutive model, including parameter identification and the validation. Finally, Section \ref{sec:conclusion} concludes this study.

\section{Methods and models}\label{sec:method}

\subsection{Systems and samples}
\label{subsec:sample}
We use the coarse-grained (CG) potential of atactic polystyrene (PS) developed by Qian et al.\,\cite{Qian2008} for MD simulations, which maps each chemical monomer to a CG bead at its center of mass. The CG potential consists of pair, bond, and angle interactions. Similar to most bottom-up CG models \cite{Reith2003,Depa2005}, this model can only represent the structural properties of its atomic counterpart in a limited temperature range, which is $400 - 500 \:\si{\kelvin}$ for the CG PS \cite{Qian2008}. Below this temperature, the structural properties is not transferable and its dynamics is accelerated due to the reduction of degree of freedom \cite{Rahimi2012}, resulting in a glass transition temperature of $\Tg\approx 170\:\si{\kelvin}$, much lower than the the values in atomistic simulations \cite{Lyulin2002} and experiments \cite{Kaliappan2005}. Therefore, this model is considered a generic model of glassy thermoplastics below its $\Tg$. The temperature transferablility could be improved by the recently proposed energy renormalization approach \cite{Xia2017,Xia2019}. However, as we focus on general properties of glassy polymers, an arbitrary traditional CG bead-spring model is sufficient for the purpose of this study. Compared to other widely-used generic bead-spring models such as the the Kremer-Grest model \cite{Kremer1990} 
and the morse model \cite{Morse1929}, the primary reason for using such a CG model is that various previous work \cite{Ries2019,Zhao2021,Zhao2024,Zhao2024a,Zhao2024b} has provided substantial useful information for choosing its parameters.

The process of preparing the MD systems has been described elsewhere in detail \cite{Zhao2024,Zhao2024a}. In brief, each MD system is generated using the self-avoiding random walk algorithm \cite{Ghanbari2011} and first equilibrated at a high temperature of \SI{590}{\kelvin} and then cooled down to the target temperature under NPT conditions. Different from our previous work \cite{Zhao2024,Zhao2024a}, we choose a much faster cooling rate of $\dT=-50\:\si{\kelvin\per\nano\second}$ and do not perform a further equilibrium simulation at the cooled target temperature. This operation significantly reduces the effects of strain softening, which increases with the state of equilibrium in glassy systems \cite{Zhao2023,vanMelick2003}, making it possible to focus on the effects of strain hardening. We consider cubic MD systems with periodic boundary conditions (PBC), comprising 500 polymer chains with 200 CG beads in each chain each.  Temperatures from $10\:\si{\kelvin}$ to $150\:\si{\kelvin}$ are considered.

All the MD simulations are performed using LAMMPS \cite{Thompson2022} with a time step size of \SI{5}{\femto\second}. The Nos\'e-Hoover thermostat and barostat \cite{Parrinello1981,Martyna1994,Shinoda2004} with coupling times of \SI{0.5}{\pico\second} and \SI{5}{\pico\second} are used, respectively.

\subsection{Uniaxial tensile simulations}
\label{subsec:deformation}

We conduct uniaxial tensile deformation in $x$-direction with constant true strain rate of $\dot{\epsilon}=1\%/\mathrm{ns}-100\%/\mathrm{ns}$ up to a maximum stretch of $\lambda_x=5$, where $\lambda_x$ is defined as $\lambda_x=l_x/L_x$ with $l_x$ and $L_{x}$ being the current and initial edge length of the MD system in the $x$-direction, respectively. The true strain rate is defined as $\dot{\epsilon}=\dot{\lambda}_x / \lambda_x$. The purpose of using true strain rate instead of engineering strain rate is to keep it consistent with the expression of deformation velocity $\llt=\dot{\FF}\FF^{-1}$ used in continuum mechanics. For convenience, we also describe the stretch using the engineering normal strain as 
\begin{align}
\varepsilon_{xx}(t)=\lambda_x(t)-1 = \exp(\dot{\epsilon}t)-1.
\label{eq:deformation_UT}
\end{align}
The surfaces in the lateral directions are subjected to NPT conditions with $p=1$ atm. It is notable that voids could form at large strains, typically above the engineering strain of $300\%$, resulting in a drop of stress with increasing deformation, which is neglected in this paper.

\subsection{Virial stress contribution}
\label{subsec:characterization}
We use the virial formulation \cite{Thompson2009} to evaluate the stress tensor, defined as
\begin{align}
\sigmaa 
&= - \frac{1}{V}\sum_{I=1}^{N_{\mathrm{A}}} m_I \vv_I \otimes \vv_I 
- \frac{1}{V}\sum_{I=1}^{N_{\mathrm{A}}} \ff_{I} \otimes \rr_{I} \label{eq:virial_stress}
\end{align}
with the volume of simulation box $V$ and the total number of CG beads $N_{\mathrm{A}}$. The first term refers to the kinematic contribution $\sigmaa^{\mathrm{ke}}$, where $m_I$ and $\vv_I$ denote the mass and velocity vector of bead $I$. The second term represents the virial stress tensor $\sigmaa^{\mathrm{virial}}$, where $\rr_I$ and $\ff_{I}$ are position vector and the resultant force vector on atom $I$. It has been derived within statistical mechanics that the virial expression~\eqref{eq:virial_stress} is equivalent to the Cauchy stress tensor with a sufficient large $N_{\mathrm{A}}$ \cite{Zimmerman2004,Subramaniyan2008}, which connects the macroscopic and microscopic mechanical responses.

To decompose the stress response between inter- and intra-chain interactions in glassy polymers, we consider the components of force $\ff_{I}$ due to pair, bond, and angle interactions as $\ff_{I}=\ff_{I}^{\mathrm{pair}}+\ff_{I}^{\mathrm{bond}}+\ff_{I}^{\mathrm{angle}}$. The virial stress is then decomposed as
\begin{align}
\sigmaa^{\mathrm{virial}} 
= \underbrace{\frac{1}{V} \sum_{I=1}^{N_{\mathrm{A}}} - \ff_{I}^{\mathrm{pair}} \otimes \rr_{I}}_{\eqqcolon \sigmaa^{\mathrm{pair}} } 
+
\underbrace{\frac{1}{V} \sum_{I=1}^{N_{\mathrm{A}}} - \ff_{I}^{\mathrm{bond}}  \otimes \rr_{I}}_{\eqqcolon \sigmaa^{\mathrm{bond}} }
+
\underbrace{\frac{1}{V} \sum_{I=1}^{N_{\mathrm{A}}} - \ff_{I}^{\mathrm{angle}}  \otimes \rr_{I}}_{\eqqcolon \sigmaa^{\mathrm{angle}} },
\label{eq:virial_compnent}
\end{align}
giving the definition of pair stress $\sigmaa^{\mathrm{pair}}$, bond stress $\sigmaa^{\mathrm{bond}}$, and angle stress $\sigmaa^{\mathrm{angle}}$.

\subsection{Microscopic properties}

We characterize the microscopic structure of local segments and global polymer chain of the MD system. For local segments, the bond length $l^{\mathrm{b}}$, bond angle $\theta^{\mathrm{a}}$ between two adjacent bond vectors, and bond orientation $P_I$ are considered. Following \cite{Hossain2010}, the bond orientation $P_I$ is defined as
\begin{align}
P_I = \frac{1}{2} \Bkt{3 \cos^2\bkt{\theta^{\mathrm{b}}_I}-1}.
\label{eq:MD_P}
\end{align}
where $\theta^{\mathrm{b}}_I$ is the angle between bond vector $I$ and the loading direction, which is the $x$ direction in this paper. These quantities are typically used in a thermodynamic average sense. For example, the averaged bond orientation is denoted as $\langle P \rangle = \sum_{I=1}^{N_{\mathrm{B}}} P_I / N_{\mathrm{B}}$ with $N_{\mathrm{B}}$ the number of bonds considered. $\langle P \rangle$ ranges from $-0.5$ to $1$, which gives the value of $-0.5$ if all bonds are orthogonal to the loading direction, while equals $1$ if they are all parallel to the loading direction. The value $\langle P \rangle =0$ means random distribution of all the bond orientations.

At the scale of polymer chains, we consider their elongation during deformation following the definition given by Hoy and Robbins \cite{Hoy2007}, described as
\begin{align}
\varepsilon_{xx}^{\mathrm{chain}}(t)=\frac{\sqrt{\langle R^{2}_{\mathrm{ee},x}(t)\rangle} - \sqrt{\langle R^{2}_{\mathrm{ee},x}(0)\rangle}}{\sqrt{\langle R^{2}_{\mathrm{ee},x}(0)\rangle}},
\label{eq:MD_strain_chain}
\end{align}
where $R_{\mathrm{ee},x}(t)$ denotes the $x$-component of the end-to-end vector of a polymer chain at time $t$ while $\langle R^{2}_{\mathrm{ee},x}(t)\rangle$ represents the thermodynamically-averaged value of its squares.

\section{Analysis of molecular dynamics results}\label{sec:MD}

\subsection{Dominating relaxation mechanisms in strain hardening}\label{subsec:MD_relaxation_mechanisms}
\begin{figure}[h!]
\centering
\includegraphics[width=0.6\textwidth]{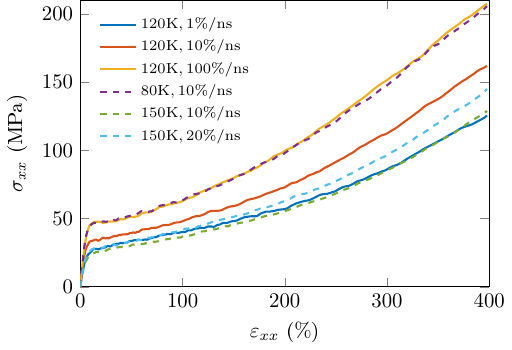}
\caption{Stress–strain curves of the MD systems in uniaxial tension simulations at different temperatures and true strain rates.}
\label{fig:fig_LS_MD_curve_stress_strain}
\end{figure}

We start with identifying the primary relaxation mechanisms of strain hardening in glassy polymers, which result in their rate- and temperature-dependent behavior as shown in Figure \ref{fig:fig_LS_MD_curve_stress_strain}. Basically, the solid curves in Figure \ref{fig:fig_LS_MD_curve_stress_strain} show rate-dependent hardening behavior at a constant temperature of $120\:\si{\kelvin}$, indicating the existence of relaxation mechanisms in the hardening region. The dashed curves in Figure \ref{fig:fig_LS_MD_curve_stress_strain} are presented to repeat the time-temperature correlation (TTC) behavior as discussed in our previous work \cite{Zhao2024} with a minimum set of simulations to compare the relaxation mechanisms at small and large strains, exhibiting different TTC behavior between small and large deformation at different temperatures.

At lower temperatures, the simulations under conditions of $[T=120\:\si{\kelvin},\dot{\epsilon}=100\%/\mathrm{ns}]$ and $[T=80\:\si{\kelvin},\dot{\epsilon}=10\%/\mathrm{ns}]$ result in approximately the same stress-strain curves at both small and large deformation, suggesting that the relaxation mechanisms in the hardening region should be the same as at small strains, i.e., before yielding. Assuming that the relaxation at the scale of polymer chains is not involved at small strains before yielding due to frozen mobility, it is reasonable to speculate that the rate-dependent strain hardening behavior at lower temperatures is also primarily determined by the relaxation of local structures, i.e., the structure composed of atoms close to each other.

In contrast, at higher temperatures close to $\Tg$, the MD system exhibits different TTC behavior at small and large strains. Specifically, the stress-strain curve at $[T=120\:\si{\kelvin},\dot{\epsilon}=1\%/\mathrm{ns}]$ coincide with the curve at $[T=150\:\si{\kelvin},\dot{\epsilon}=20\%/\mathrm{ns}]$ at strains below $80\%$ but approximately coincide with the curve at $[T=150\:\si{\kelvin},\dot{\epsilon}=10\%/\mathrm{ns}]$ at strains above $300\%$. This different TTC behavior at small and large strains means that the dominating relaxation mechanisms in the hardening region at higher temperatures is not only the relaxation of local structures. Most likely, the relaxation of polymer chains such as disentanglement start contributing to the rate- and temperature-dependence of strain hardening at this temperature. As discussed in Section \ref{sec:introduction}, this conjecture can well interpret the contradictory results of MD simulations for different glassy polymer models reported in \cite{Jatin2014,Zhu2022}, where the entanglement length $\Ne$ remains approximately constant in \cite{Jatin2014} but evolves with increasing strain in the hardening region in \cite{Zhu2022}.

\begin{figure}[h!]
\centering
\includegraphics[width=0.6\textwidth]{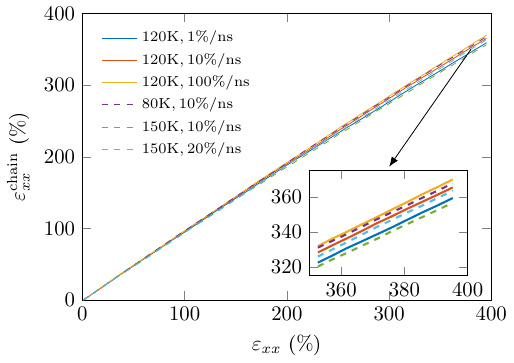}
\caption{The averaged stretch of polymer chains as defined in Equation \eqref{eq:MD_strain_chain} as a function of applied engineering strain at different temperatures and true strain rates. }
\label{fig:fig_LS_MD_curve_chain_stretch}
\end{figure}

To observe the effects of disentanglement at higher temperatures, we follow~\cite{Hoy2007} to compare the stretch of polymer chains under different conditions during deformation as depicted in Figure \ref{fig:fig_LS_MD_curve_chain_stretch}, where the averaged stretch of polymer chains is scaled in the form of strain as defined in Equation \eqref{eq:MD_strain_chain}. The chain stretch $\varepsilon_{xx}^{\mathrm{chain}}$ can follow the macroscopic strain $\varepsilon_{xx}$ with increasing deformation, indicating that the effects of relaxation at the scale of polymer chains play a minor role in strain hardening. Instead, polymer chains are stretched approximately affine to the macroscopic deformation. It is the relaxation of local structures such as chain segments that dominates the rate- and temperature-dependent strain hardening. At different temperatures and strain rates, the chain stretch $\varepsilon_{xx}^{\mathrm{chain}}$ coincide at small strains and exhibit slight deviations at large strains, where the chain stretch at higher temperatures and lower strain rates becomes smaller than that at lower temperatures and higher strain rates, indicating that disentanglement happens under these conditions, but only resulting in a small contribution to the relaxation at the scale of polymer chains. This result is consistent with the observation in \cite{Hoy2007} and the results shown in Figure \ref{fig:fig_LS_MD_curve_stress_strain}. In summary, the stress increase in strain hardening is caused by chain stretch, primarily mediated by relaxation of local structures at temperatures below $\Tg$, and slightly influenced by the relaxation at larger length scales such as disentanglement at the temperatures close to $\Tg$.

\subsection{Primary contributions to strain hardening}\label{subsec:MD_primary_contributions}
\begin{figure}[h!]
\centering
\begin{subfigure}[t]{0.48\textwidth}
\centering
\caption{}
\includegraphics[width=0.92\textwidth]{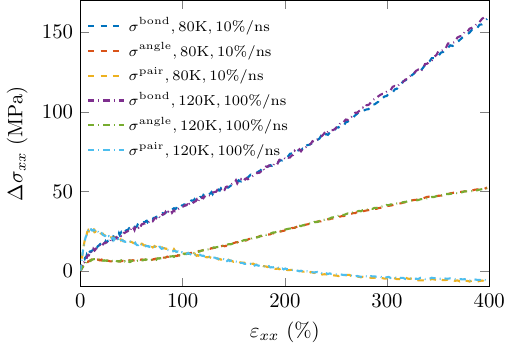}
\end{subfigure}%
\hfill
\begin{subfigure}[t]{0.48\textwidth}
\centering
\caption{}
\includegraphics[width=0.92\textwidth]{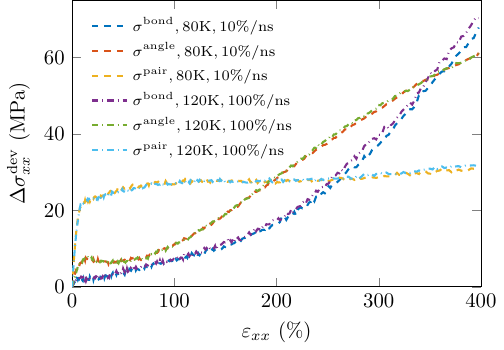}
\end{subfigure}%
\caption{The virial stress contributions as defined in Equation \eqref{eq:virial_compnent} as a function of engineering strain in the form of (a) the normal component in the loading direction and (b) the deviatoric part. The $y$ axis $\Delta \sigma$ denoted the in crease of stress compared to its initial value as $\Delta \sigma (t) = \sigma (t)-\sigma (0)$ at time $t$.}
\label{fig:fig_LS_MD_curve_stress_strain_virial}
\end{figure}

We further attempt to identify which local structure is the main resource to strain hardening in glassy polymers. To this end, we consider the virial components of stress responses during deformation as defined in Equation \eqref{eq:virial_compnent}. We focus on the simulations at lower temperatures to avoid the influence of disentanglement. Figure \ref{fig:fig_LS_MD_curve_stress_strain_virial}(a) depicts the $x$ components of bond, angle, and pair stresses during deformation at the conditions of $[T=120\:\si{\kelvin},\dot{\epsilon}=100\%/\mathrm{ns}]$ and $[T=80\:\si{\kelvin},\dot{\epsilon}=10\%/\mathrm{ns}]$, which is adopted from the simulation sets resulting in coinciding stress-strain curves in Figure~\ref{fig:fig_LS_MD_curve_stress_strain}. It is evident in Figure \ref{fig:fig_LS_MD_curve_stress_strain_virial}(a) that the virial components also coincide in these simulations, suggesting that the dominating relaxation mechanism also has very similar influence on each virial component. In Figure \ref{fig:fig_LS_MD_curve_stress_strain_virial}(a), the pair stress initially increases to a yield point and then decreases continuously, suggesting that the pair interaction is not the primary contribution to strain hardening. In contrast, the strain hardening should be attributed to the bond and angle stresses as they increase in the hardening region. 

To explain the drop of the $x$ component of the pair stress in Figure \ref{fig:fig_LS_MD_curve_stress_strain_virial}(a), we further consider the deviatoric part of the stress, which is defined as $\sigma_{xx}^{\mathrm{dev}}=[2\sigma_{xx}-\sigma_{yy}-\sigma_{zz}]/3$ in uniaxial deformation, as depicted in Figure \ref{fig:fig_LS_MD_curve_stress_strain_virial}(b). Here, the deviatoric pair stress remains approximately constant after yielding, indicating that pair interactions also has no negative contributions to strain hardening. As explained in our previous work \cite{Zhao2024a}, the decrease of pair stress in the loading direction could be attributed to bond orientation, which squeezes the space in the loading direction but releases more space along the lateral directions for pair interactions. Similar to their $x$ components, the deviatoric part of bond and angle stresses also increases in the hardening region, justifying that the strain hardening is caused by bond and angle interactions in the deviatiric point of view.

\begin{figure}[h!]
\centering
\begin{subfigure}[t]{0.48\textwidth}
\centering
\caption{}
\includegraphics[width=0.92\textwidth]{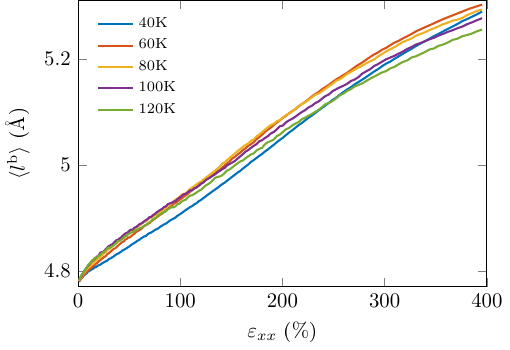}
\end{subfigure}%
\hfill
\begin{subfigure}[t]{0.48\textwidth}
\centering
\caption{}
\includegraphics[width=0.92\textwidth]{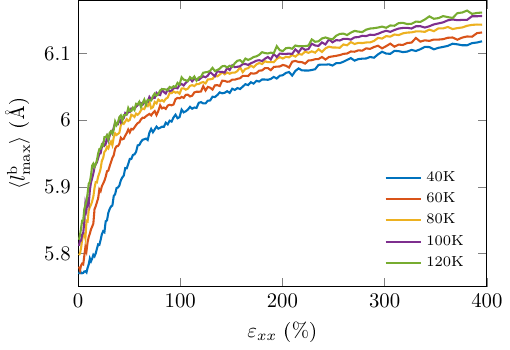}
\end{subfigure}%
\\
\begin{subfigure}[t]{0.48\textwidth}
\centering
\caption{}
\includegraphics[width=0.92\textwidth]{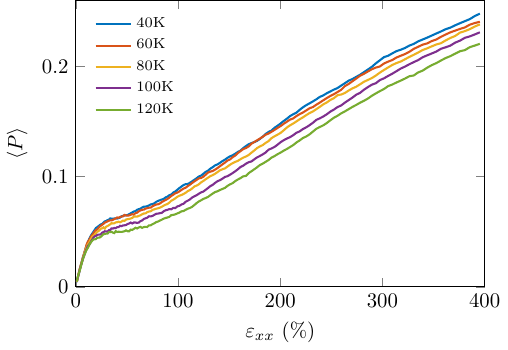}
\end{subfigure}%
\hfill
\begin{subfigure}[t]{0.48\textwidth}
\centering
\caption{}
\includegraphics[width=0.92\textwidth]{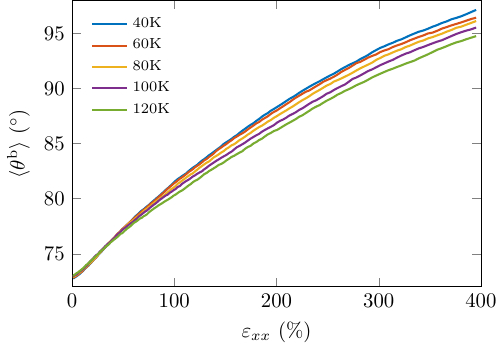}
\end{subfigure}%
\caption{The averaged microscopic quantities of the MD system along deformation with a strain rate of $100\%/\mathrm{ns}$ at different temperatures: (a) averaged bond length, (b) averaged value of the largest bond length in each polymer chain, (c) averaged bond orientation as defined in Equation \eqref{eq:MD_P}, and (d) averaged bond angle.}
\label{fig:fig_LS_MD_curve_micro_deformation}
\end{figure}

To further explore the exact source contributing to the stress response in strain hardening, we compare the evolution of different quantities representing the microscopic structures in  Figure \ref{fig:fig_LS_MD_curve_micro_deformation}. Figure \ref{fig:fig_LS_MD_curve_micro_deformation}(a) presents the evolution of the averaged bond length $\langle l^{\mathrm{b}} \rangle$ during deformation at a strain rate of $\dot{\epsilon}=100\%/\mathrm{ns}$ at different temperatures. A nonmonotonic dependence of $\langle l^{\mathrm{b}} \rangle$ on the temperature suggests that $\langle l^{\mathrm{b}} \rangle$ is not an appropriate variable to describe the microscopic structure associated with strain hardening. Instead, we use the averaged value of the maximum bond length in each polymer chain $\langle l^{\mathrm{b}}_{\max} \rangle$ to characterize the microscopic structure. As shown in Figure \ref{fig:fig_LS_MD_curve_micro_deformation}(b), the variable $\langle l^{\mathrm{b}}_{\max} \rangle$ monotonically increases with decreasing temperatures in the hardening region. Furthermore, the transition similar to yielding can be observed in the curves in Figure \ref{fig:fig_LS_MD_curve_micro_deformation}(b). These features indicate that $\langle l^{\mathrm{b}}_{\max} \rangle$ could be a good candidate to characterize the microscopic structure associated with strain hardening. Moreover, it is reasonable to imagine that the traction of an elongated chain should be balanced by its longest bond if all bonds are connected in series. We denote this bond as load-bearing bond. We further denote the local deformation associated with the load-bearing bonds as local load-bearing deformation.

Figure \ref{fig:fig_LS_MD_curve_micro_deformation}(c-d) presents the evolution of averaged chain orientation $\langle P \rangle$ and bond angle $\langle \theta^{\mathrm{b}} \rangle$ during deformation, respectively, which increases with decreasing temperature in the hardening region. These two variables can represent the relaxation of the chain segments, where a slower increase during deformation means a stronger relaxation, caused by higher temperatures. This relaxation accommodates the chain stretch to mediate the stress increase.

\subsection{The quantitative effects load-bearing bond}\label{subsec:MD_quantitative_effects_lb_max}
Following the finding that the stress in the hardening region is primarily contributed by the longest bond in each chain, i.e., the load-bearing bonds, we explore their quantitative effects on strain hardening in this section.

\begin{figure}[h!]
\centering
\begin{subfigure}[t]{0.48\textwidth}
\centering
\caption{}
\includegraphics[width=0.92\textwidth]{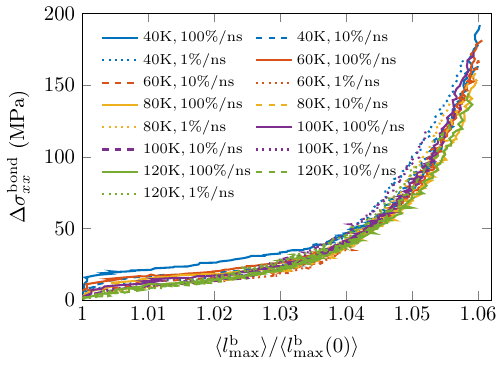}
\end{subfigure}%
\hfill
\begin{subfigure}[t]{0.48\textwidth}
\centering
\caption{}
\includegraphics[width=0.92\textwidth]{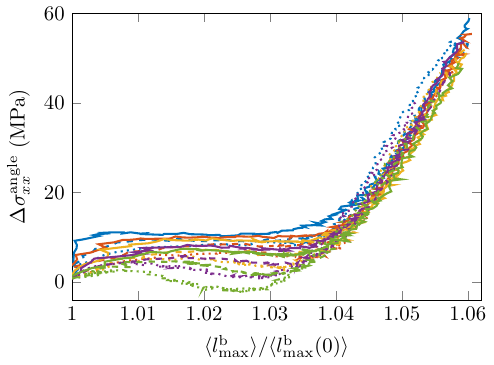}
\end{subfigure}%
\caption{The response of (a) bond stress and (b) angle stress as a function of the stretch of the averaged value of the longest bonds in each chain at different temperatures and strain rates.}
\label{fig:fig_LS_MD_curve_micro_local_virial}
\end{figure}

In Figure \ref{fig:fig_LS_MD_curve_micro_local_virial}(a-b), the evolution of bond stress and angle stress are plotted against the change of the averaged lengths of load-bearing bonds, which are represented in the form of relative stretch $\langle l^{\mathrm{b}}_{\max} \rangle / \langle l^{\mathrm{b}}_{\max}(0) \rangle$. Here, $\langle l^{\mathrm{b}}_{\max}(0) \rangle \rangle$ denotes the value of $\langle l^{\mathrm{b}}_{\max} \rangle$ before deformation. It is evident that both the bond and angle stresses approximately coincide at larger stretch of load-bearing bonds, corresponding to the hardening region. This result suggest that the bond and angle stresses are dominated by the stretch of load-bearing bonds, independent of the relaxation of chain segments, for example, characterized by the evolution of $\langle P \rangle$ and $\langle \theta^{\mathrm{b}} \rangle$.

\begin{figure}[h!]
\centering
\includegraphics[width=0.7\textwidth]{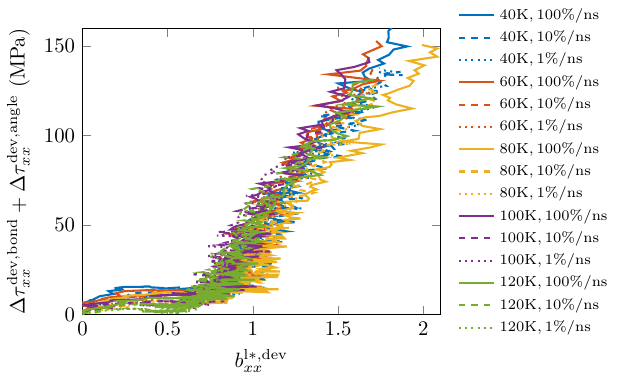}
\caption{Deviatoric part of the bonded stress (sum of the bond and angle stress) as a function of the deviatoric part of the left Cauchy-Green deformation tensor for the load-bearing stretch in the loading direction as defined in Equation \eqref{eq:deform_bl_dev}.}
\label{fig:fig_LS_MD_curve_micro_local_dev}
\end{figure}

We relate the stretch of load-bearing bonds to continuum mechanics by representing them using a local deformation gradient, denoted as $\FFl$. In uniaxial deformation, $\FFl$ comprises only the diagonal components as $F^{\mathrm{l}}_{ii}=\langle l^{\mathrm{b}}_{\max,i} \rangle /  \langle l^{\mathrm{b}}_{\max,i}(0)  \rangle$ with $i=x,y,z$, where $l^{\mathrm{b}}_{\max,i}$ denotes the projection of $l^{\mathrm{b}}_{\max}$ in direction $i$. Then we have the left Cauchy-Green tensor $\bb^{\mathrm{l}}=\FFl\FF^{\mathrm{lT}}$ with $\FF^{\mathrm{lT}}=\Bkt{\FF^{\mathrm{l}}}^{\mathrm{T}}$. Its isochoric component of the deviatoric part is given by
\begin{align}
\bb^{\mathrm{l}*,\mathrm{dev}}= J^{\mathrm{l}-\frac{2}{3}} \, \bb^{\mathrm{l}} - \frac{1}{3} \unittensor \, \tr \bkt{J^{\mathrm{l}-\frac{2}{3}}  \,\bb^{\mathrm{l}} }
\label{eq:deform_bl_dev}
\end{align}
with the Jacobian determinant $J^{\mathrm{l}}=\det\,\FFl$ and the second-order unit tensor in the deformed configuration $\unittensor$. Figure \ref{fig:fig_LS_MD_curve_micro_local_dev} depicts the deviatoric part of the bonded Kirchhoff stresses with respect to the evolution of the $x$ components of $\bb^{\mathrm{l}*,\mathrm{dev}}$, showing that the bonded deviatoric Kirchhoff stresses are almost proportional to $b_{xx}^{\mathrm{l}*,\mathrm{dev}}$ in the hardening region, independent of temperatures and strain rates. This relation can be formulated as
\begin{align}
\tauu^{\mathrm{bonded},\mathrm{dev}} = \mu^{\mathrm{b}} \dev\bkt{\bb^{\mathrm{l}*}},
\label{eq:NH_MD_bond}
\end{align}
which is equivalent to the formulation of Neo-Hookean (NH) model utilizing $\tauu=J\sigmaa$ with $J=\det \FF$. Here, the sum of the bond and angle stress is denoted as bonded stress as they exhibit similar dependence on the local load-bearing stretch as shown in Figure \ref{fig:fig_LS_MD_curve_micro_local_virial}. Through fitting the curves in the hardening region in Figure \ref{fig:fig_LS_MD_curve_micro_local_dev}, the value of the bond shear modulus is obtained as $\mu^{\mathrm{b}}=131\:\si{\mega\pascal}$ is obtained. Based on the relation of the bonded stress and load-bearing stretch in the hardening region, we assume that $\mu^{\mathrm{b}}$ is the intrinsic hardening modulus.

In summary, the results in this section suggests that the strain hardening is caused by the stretch of chain segments, primarily mediated by the relaxation of local structures at temperatures far below $\Tg$ and also slightly influenced by disentanglement at temperatures close to $\Tg$. The stress increase during strain hardening is primarily contributed by bonded interactions, i.e., the bond stress and angle stress, while the pair stress only has negligible influence in the hardening region. The bonded stress is mainly contributed by the stretch of local load-bearing bond, which can be related to continuum mechanics by defining a local load-bearing deformation gradient $\FFl$. Such a definition results in a relation between the bonded stress and $\FFl$, independent of temperatures and strain rates, leading to the NH constitutive model characterized by the bond shear modulus $\mu^{\mathrm{b}}$.

\section{Decomposition of the deformation gradient}\label{sec:decomposition}

\begin{figure}[h!]
\centering
\includegraphics[width=0.4\textwidth]{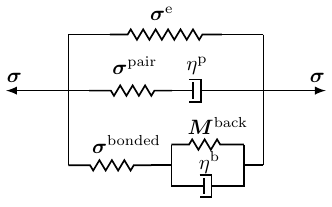}
\caption{Rheological representation of the constitutive model.}
\label{fig:fig_constitutive_model}
\end{figure}

Similar to the decomposition of the virial stress, we assume that the total stress in continuum mechanics comprises the contribution from pair, bonded and an elastic term as represented by the rheological model in Figure \ref{fig:fig_constitutive_model}:
\begin{align}
\sigmaa=\sigmaa^{\mathrm{pair}}+\sigmaa^{\mathrm{bonded}}+\sigmaa^{\mathrm{e}},
\end{align}
where $\sigmaa^{\mathrm{bonded}}$ is the stress due to bond and angle interactions. The elastic term $\sigmaa^{\mathrm{e}}$ is defined for completeness of the constitutive model, which might be mainly attributed to volumetric deformation. We assume these stress terms are independent and our focus in this section is on the analysis of the bonded branch.

\subsection{Decomposition of deformation gradient}
Based on the results in Section \ref{subsec:MD_quantitative_effects_lb_max} that the bonded stress is merely contributed by the local load-bearing deformation gradient $\FFl$, it is reasonable to assume a decomposition of the total deformation into two parts as $\FF=\FFl\FFr$, where $\FFr$ is denoted as the resistance part of the deformation gradient. This decomposition corresponds to the illustration of the bonded branch in Figure \ref{fig:fig_constitutive_model}, where $\FFl$ contributes to $\sigmaa^{\mathrm{bonded}}$ while the evolution of $\FFr$ is driven by $\sigmaa^{\mathrm{bonded}}$. As the part of chain segments corresponding to $\FFr$ also has stress caused by bonded interactions, which resists itself to be stretched, the resultant driving force should be the difference between $\sigmaa^{\mathrm{bonded}}$ and the resistant stress. In convention, we denote this resistance stress as back stress. The driving stress leads to the evolution of $\FFr$, balanced by the friction due to sliding of polymer segments. The friction is considered as rate-dependent and characterized by the viscosity $\eta^{\mathrm{b}}$. While in mathematical representation, our decomposition $\FF=\FFl\FFr$ is the same as the Lee-Kröner decomposition $\FF=\FFe\FFp$ \cite{Kroener1959,Lee1969}, the physical meaning becomes clear in the decomposition $\FF=\FFl\FFr$ for glassy polymers.

The velocity gradient of the resistant part is defined as 
\begin{align*}
\llt^{\mathrm{r}}=\dot{\FF}^{\mathrm{r}}\FF^{\mathrm{r}-1},
\end{align*}
and the corresponding rate of deformation tensor is given by $\DD^{\mathrm{r}}=[\llt^{\mathrm{r}}+\llt^{\mathrm{rT}}]/2$. Assuming isochoric resistance deformation $\FFr$ analogous to most study of inelastic flow in glassy polymers, i.e., $\det\,\FFr=1$, the relation between $\DD^{\mathrm{r}}$ and the driving effective Mandel stress $\MM^{\mathrm{eff,dev}}=\MM^{\mathrm{bonded,dev}}-\MM^{\mathrm{back,dev}}$ is given by
\begin{align}
\DD^{\mathrm{r}} = \frac{\MM^{\mathrm{eff,dev}}}{2 \eta^{\mathrm{b}}},
\label{eq:flow_resistance_Meff}
\end{align}
where $\MM^{\mathrm{bonded,dev}}=\FF^{\mathrm{lT}}\, \tauu^{\mathrm{bonded,dev}}\, \FF^{\mathrm{l-T}}$ is the pull-back of the bonded Kirchhoff stress $\tauu^{\mathrm{bonded,dev}}=J\sigmaa^{\mathrm{bonded,dev}}$ to the intermediate configuration. The superscript ``dev'' denotes the deviatoric part of the stress tensor.

\subsection{Deformation of the resistance stretch}
\begin{figure}[h!]
\centering
\begin{subfigure}[t]{0.48\textwidth}
\centering
\caption{}
\includegraphics[width=0.92\textwidth]{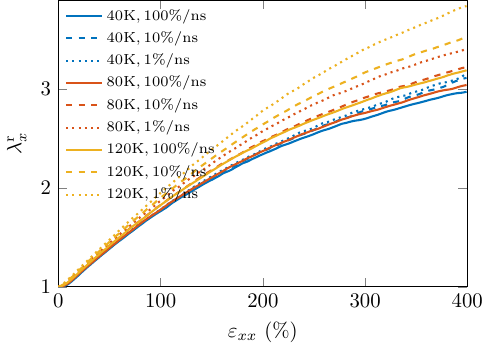}
\end{subfigure}%
\hfill
\begin{subfigure}[t]{0.48\textwidth}
\centering
\caption{}
\includegraphics[width=0.92\textwidth]{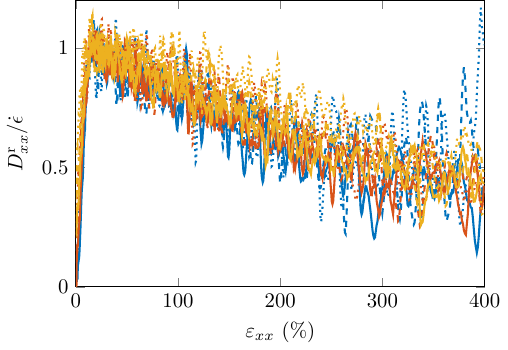}
\end{subfigure}%
\caption{Estimated (a) stretch and (b) rate of deformation of the resistance part of the deformation gradient $\FFr$ in the loading direction. In (b), the rate of deformation is scaled by the prescribed true strain rate.}
\label{fig:fig_LS_MD_curve_decomposition_resistance}
\end{figure}

Based on the results obtained in Section \ref{sec:MD}, the isochoric part of the load-bearing deformation gradient $\FF^{\mathrm{l}*}$ can be recovered using the relation \eqref{eq:NH_MD_bond} with given bond shear modulus $\mub$ and bonded stress in MD simulations. Then, it is straightforward to calculate the components of $\FFr$ and its rate $\dot{\FF}^{\mathrm{r}}$ using the decomposition $\FF=\FFl\FFr$. Figure \ref{fig:fig_LS_MD_curve_decomposition_resistance}(a) presents the evolution of the stretch of the resistance part in the loading direction $\lambda^{\mathrm{r}}_x$ at various temperatures and strain rates, where $\lambda^{\mathrm{r}}_x$ is the $x$ component in $\FFr$. An evident rate- and temperature-dependence of $\lambda^{\mathrm{r}}_x$ can be observed. The resistance stretch in simulations at higher temperature and lower strain rate shows more tendency to follow the total deformation of the systems. Figure \ref{fig:fig_LS_MD_curve_decomposition_resistance}(b) illustrates the $x$ component of $\DD^{\mathrm{r}}$ scaled by the loading rate during deformation at the same temperature and rate conditions as in Figure \ref{fig:fig_LS_MD_curve_decomposition_resistance}(a). It is striking that all the scaled curves approximately coincide during the whole deformation process, suggesting that the flow rate of the resistance stretch is independent of temperature. Instead, it can follow the stretch rate of the whole system $\dot{\epsilon}$ in a scaled point of view. A yield point can be evidently observed at the highest point of the curves, where $D^{\mathrm{r}}_{xx} \approx \dot{\epsilon}$, indicating that at the yield point only the resistance part is deformed while the load-bearing stretch ${\lambda}^{\mathrm{l}}_x$ is fixed. After the yield point, the decrease of $D^{\mathrm{r}}_{xx}$ corresponds to the increase of the load-bearing stretch $\dot{\lambda}^{\mathrm{l}}_x$, leading to the increase of bonded stress in the hardening region.

\subsection{Yielding at small deformation}\label{subsec:decomposition_yielding}
\begin{figure}[h!]
\centering
\includegraphics[width=0.6\textwidth]{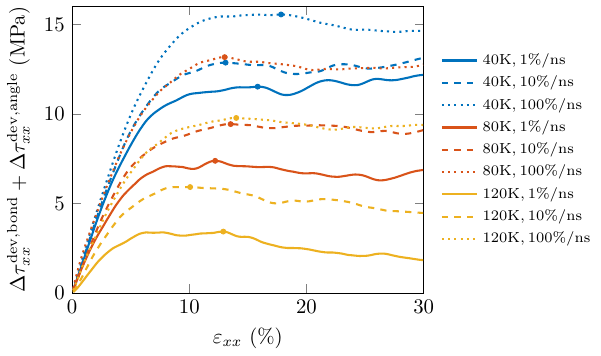}
\caption{The response of bonded stress of the MD system during deformation at small strains marked by the yield points.}
\label{fig:fig_LS_MD_curve_decomposition_vb_small_strain}
\end{figure}

To analyze the stress arising from the flow of the resistance deformation, we consider the yield behavior of the bonded stress at small strains. Figure \ref{fig:fig_LS_MD_curve_decomposition_vb_small_strain} shows the stress-strain curves of the bonded contributions in MD simulations at strains below $30\%$, where the yield points are marked by solid points. The yield stress is estimated as the highest point on the curves at strains below $20\%$. As the bonded stress flattens after yielding, the errors of the yield stresses resulting from this estimation method are acceptable.

The estimated yield stresses are used to identify the parameters in the viscosity model $\eta^{\mathrm{b}}$ in Equation \eqref{eq:flow_resistance_Meff}. Here, we consider the Eyring model
\begin{align}
\eta^{\mathrm{b}} 
&= \eta^{\mathrm{b}}_0(T) \frac{Q m^{\mathrm{eq}}}{T} \Bkt{\sinh\bkt{\frac{Q m^{\mathrm{eq}}}{T}}}^{-1} \nonumber\\
&\approx \eta^{\mathrm{b}}_0(T) \frac{2Q m^{\mathrm{eq}}}{T} \Bkt{\exp\bkt{\frac{Q m^{\mathrm{eq}}}{T}}}^{-1}
\label{eq:Eyring_model}
\end{align}
with the specific activation volume $Q$ and equivalent stress $m^{\mathrm{eq}}=\sqrt{\MM^{\mathrm{eff,dev}} : \MM^{\mathrm{eff,dev}}/2}$. The variable $\eta^{\mathrm{b}}_0(T)$ is a temperature-dependent viscosity at zero stress, denoted as static viscosity. The approximation sign is Equation \eqref{eq:Eyring_model} is valid if $m^{\mathrm{eq}}$ is sufficiently large. In uniaxial deformation alone $x$-axis, $m^{\mathrm{eq}}=\sqrt{3} |M^{\mathrm{eff,dev}}_{xx}|/2=\sqrt{3} |\tau^{\mathrm{eff,dev}}_{xx}|/2$.

\begin{figure}[h!]
\centering
\begin{subfigure}[t]{0.48\textwidth}
\centering
\caption{}
\includegraphics[width=0.92\textwidth]{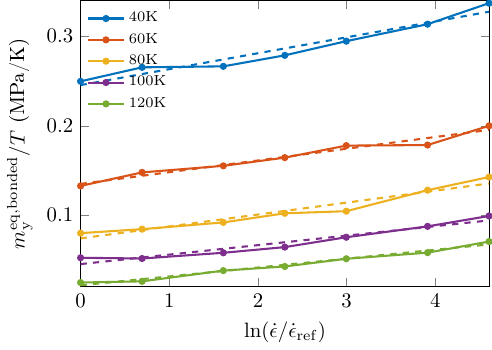}
\end{subfigure}%
\hfill
\begin{subfigure}[t]{0.48\textwidth}
\centering
\caption{}
\includegraphics[width=0.92\textwidth]{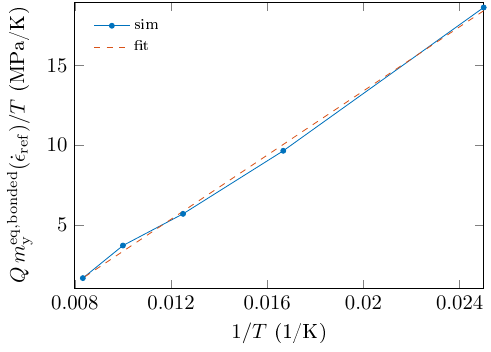}
\end{subfigure}%
\caption{(a) Bonded yield stress as a function of scaled strain rates at different temperatures with $\dot{\epsilon}_{\mathrm{ref}}=1\%/\mathrm{ns}$ and their linear fitting (dashed). (b) The quantity of $Q\,m^{\mathrm{eq,bonded}}_{\mathrm{y}}/{T}$ at the strain rate of $\dot{\epsilon}_{\mathrm{ref}}$ as a function of $1/T$ from simulations (solid curve) and the linear fit (dashed curve).}
\label{fig:fig_LS_MD_curve_decomposition_yield_stress_rate}
\end{figure}

Assuming that before yielding there is no back stress, we have $\MM^{\mathrm{eff,dev}}=\MM^{\mathrm{bonded,dev}}$ at small strains. Utilizing the approximation at the yield point that $D^{\mathrm{r}}_{xx, \mathrm{y}}\approx\dot{\epsilon}$ as demonstrated in Figure \ref{fig:fig_LS_MD_curve_decomposition_resistance}(b) and combining Equations \eqref{eq:flow_resistance_Meff} and \eqref{eq:Eyring_model}, we have 
\begin{align}
M^{\mathrm{bonded,dev}}_{xx,\mathrm{y}} = \eta^{\mathrm{b}}_0(T) \frac{2\sqrt{3} Q |M^{\mathrm{bonded,dev}}_{xx,\mathrm{y}}|}{T} \Bkt{\exp\bkt{\frac{Q m^{\mathrm{eq}}_{\mathrm{y}}}{T}}}^{-1} \dot{\epsilon},
\label{eq:flow_resistance_component}
\end{align}
where the subscript ``y'' denotes the stress at the yield point. Given $M^{\mathrm{bonded,dev}}_{xx,\mathrm{y}}=\tau^{\mathrm{bonded,dev}}_{xx,\mathrm{y}}>0$ in uniaxial tension as evident in the MD simulation results, Equation \eqref{eq:flow_resistance_component} can be transformed into
\begin{align}
\frac{m^{\mathrm{eq}}_{\mathrm{y}}}{T} = \frac{1}{Q}\ln\bkt{\frac{2\sqrt{3} \,\eta^{\mathrm{b}}_0(T) Q\dot{\epsilon}_{\mathrm{ref}}}{T}} + \frac{1}{Q}\ln\bkt{\frac{\dot{\epsilon}}{\dot{\epsilon}_{\mathrm{ref}}}},
\label{eq:flow_resistance_identify_Q}
\end{align}
where the variable $\dot{\epsilon}_{\mathrm{ref}}$ is an arbitrary constant reference strain rate. The dependence of ${m^{\mathrm{eq}}_{\mathrm{y}}}/{T}$ on $\ln(\dot{\epsilon} / \dot{\epsilon}_{\mathrm{ref}})$ at different temperatures and strain rates are plotted in Figure \ref{fig:fig_LS_MD_curve_decomposition_yield_stress_rate}(a), where $\dot{\epsilon}_{\mathrm{ref}}=1\%/\mathrm{ns}$ is used. The specific activation volume $Q$ can be fitted, giving an averaged value of $Q=80.54 \:\si{\kelvin\per\mega\pascal}$.

We further identify the temperature dependent static viscosity $\eta^{\mathrm{b}}_0(T)$ by introducing a shift factor $a(T)$ that $\eta^{\mathrm{b}}_0(T)=a(T)\eta^{\mathrm{b}}_{\mathrm{g}0}$ with the static viscosity $\eta^{\mathrm{b}}_{\mathrm{g}0}$ at $\Tg$. Inserting this relation into Equation \eqref{eq:flow_resistance_identify_Q}, the shift factor $a(T)$ can be expressed as
\begin{align}
\ln(a(T))=
 \frac{Q m^{\mathrm{eq}}_{\mathrm{y}}}{T} - \ln\bkt{\frac{\Tg}{T}}
 -\ln\bkt{\frac{2\sqrt{3} \,\eta^{\mathrm{b}}_{\mathrm{g}0}Q\dot{\epsilon}_{\mathrm{ref}}}{\Tg}}
 - \ln\bkt{\frac{\dot{\epsilon}}{\dot{\epsilon}_{\mathrm{ref}}}},
\label{eq:flow_resistance_aT}
\end{align}
where the last two terms on the right hand side are independent of temperature. Using Arrhenius function that $a(T)= \exp\bkt{A/T-A/\Tg}$, we can identify the constant $A$ by fitting the curves of $\Bkt{{Q\tau^{\mathrm{eq}}_{\mathrm{y}}}/{T} - \ln\bkt{{\Tg}/{T}}}$ with respect to $1/T$ with a linear function as shown in Figure \ref{fig:fig_LS_MD_curve_decomposition_yield_stress_rate}(b). We obtain $A=1008\:\si{\kelvin}$ and the relation
\begin{align}
&\quad Q\frac{m^{\mathrm{eq}}_{\mathrm{y}}}{T} - \ln\bkt{\frac{\Tg}{T} }
= \frac{A}{T}+B 
\label{eq:flow_resistance_aT_fit}
\end{align}
with $B=-6.749$. Comparing Equations \eqref{eq:flow_resistance_aT} and \eqref{eq:flow_resistance_aT_fit}, the static viscosity at $\Tg$ can be expressed as
\begin{align}
\eta^{\mathrm{b}}_{\mathrm{g}0} &= \frac{\Tg}{2\sqrt{3}Q\dot{\epsilon}_{\mathrm{ref}}}\exp\bkt{B+\frac{A}{\Tg}},
\end{align}
which results in $\eta_{\mathrm{g}0}=26.86 \:\si{\mega\pascal\nano\second}=2.686\times10^{-8}\:\si{\mega\pascal\second}$.

It is notable that the values of $Q$, $A$, and $\eta^{\mathrm{b}}_{\mathrm{g}0}$ are calculated based on the following assumptions at the yield point: i) the back stress has negligible contributions to the flow of the resistance deformation; ii) the flow rate of the resistance deformation is the same as the prescribed strain rate; iii) The Eyring model and Arrhenius function are applicable for the MD systems at temperatures below $\Tg$ at small strains around the yield point. While the results derived from the last two assumptions seem to be reasonable as evident from the MD simulation results shown in Figures \ref{fig:fig_LS_MD_curve_decomposition_resistance}(b) and \ref{fig:fig_LS_MD_curve_decomposition_yield_stress_rate}(a-b), the first assumption could result in oversimplification and errors of the estimated values. The values of these quantities will be further checked in constitutive models discussed in Section \ref{sec:constitutive}.

\subsection{Back stress}
\begin{figure}[h!]
\centering
\begin{subfigure}[t]{0.48\textwidth}
\centering
\caption{}
\includegraphics[width=0.92\textwidth]{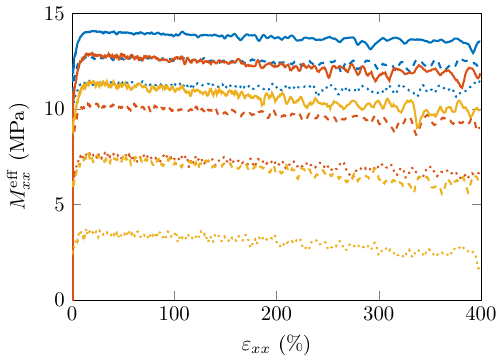}
\end{subfigure}%
\hfill
\begin{subfigure}[t]{0.48\textwidth}
\centering
\caption{}
\includegraphics[width=0.92\textwidth]{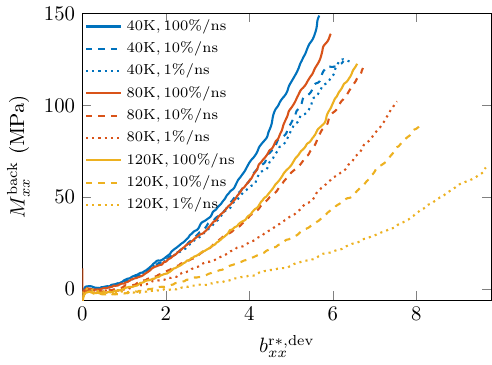}
\end{subfigure}%
\caption{Estimated (a) effective Mandel stress and (b) back stress in the deviatoric form.}
\label{fig:fig_LS_MD_curve_decomposition_stress_eff_back}
\end{figure}

The remaining component in the bonded branch in Figure \ref{fig:fig_constitutive_model} is the expression of the back stress. We write the $x$ component in Equation \eqref{eq:flow_resistance_Meff} as 
\begin{align}
M^{\mathrm{eff,dev}}_{xx}=2\eta^{\mathrm{b}} D^{\mathrm{r}}_{xx}
= 
\eta^{\mathrm{b}}_{\mathrm{g0}}a(T) \frac{2Q m^{\mathrm{eq}}}{T} \Bkt{\sinh\bkt{\frac{Q m^{\mathrm{eq}}}{T}}}^{-1} D^{\mathrm{r}}_{xx}.
\end{align}
In uniaxial tension, we can solve this equation utilizing the relation $m^{\mathrm{eq}}={\sqrt{3}} M^{\mathrm{eff,dev}}_{xx} /2$, which gives 
\begin{align}
M^{\mathrm{eff,dev}}_{xx} = \frac{2}{\sqrt{3}} \frac{T}{Q}\sinh^{-1}\bkt{ \frac{\sqrt{3}\eta^{\mathrm{b}}_{\mathrm{g0}}\,a(T)Q D^{\mathrm{r}}_{xx}}{T}}.
\end{align}
We plot the results of the estimated $M^{\mathrm{eff,dev}}_{xx}$ as a function of strain at different temperatures and strain rates in Figure \ref{fig:fig_LS_MD_curve_decomposition_stress_eff_back}(a) and found that the effective Mandel stress flattens in the hardening region. While this is based on the assumption that the viscosity $\eta^{\mathrm{b}}$ is independent of the shape of the system, these curves cannot provide quantitative results for the analysis of the microscopic mechanisms of strain hardening bur only serve to provide some insights for the subsequent constitutive modeling.

Subtracting the estimated effective stress from the total bonded Mandel stress, we obtain the estimated back stress $M^{\mathrm{back}}_{xx}$. We plot them as a function of the deviatoric part of the left Cauchy green tensor for the resistance deformation $b^{\mathrm{r*,dev}}_{xx}$ in Figure \ref{fig:fig_LS_MD_curve_decomposition_stress_eff_back}(b). The definition of $b^{\mathrm{r*,dev}}_{xx}$ is similar to their counterpart $b^{\mathrm{l*,dev}}_{xx}$ given in Equation \eqref{eq:deform_bl_dev}. The rate- and temperature-dependence of the curves in Figure \ref{fig:fig_LS_MD_curve_decomposition_stress_eff_back}(b) indicates that the back stress $\MM^{\mathrm{back}}$ is not a unary function of $\FFr$. Instead, it is appropriate to introduce an internal variable to construct the function of the back stress, such as the widely used orientation tensor $\AA$ \cite{Anand2009}. Furthermore, the shape of the curves suggests that the relation between $\MM^{\mathrm{back}}$ and $\AA$ should not be the NH model as the slopes increases with deformation. A Langevin-based model could be considered as used in \cite{Xiao2019}.

\section{Constitutive modeling}\label{sec:constitutive}
In this section, we synthesize the results obtained in previous sections in a constitutive model as schematically illustrated in Figure \ref{fig:fig_constitutive_model} to validate the mechanism of local load-bearing deformation that primarily induces strain hardening in glassy polymers. In the pair branch, the conventional Lee-Kröner decomposition of the deformation gradient $\FF=\FFe\FFp$ \cite{Kroener1959,Lee1969} is used while in the bonded branch, the decomposition $\FF=\FFl\FFr$ discussed in Section \ref{sec:MD} is assumed. We start with deriving the mathematical expression in Section \ref{subsec:CM_TD} and \ref{subsec:CM_individual_models}, followed by parameter identification and validation in Section \ref{subsec:CM_PI} and \ref{subsec:CM_validation}, respectively.

\subsection{Thermodynamics}\label{subsec:CM_TD}
The requirement of the constitutive model for thermodynamic consistency is firstly derived in this subsection. Different from the conventional Coleman-Noll procedure \cite{Coleman1963,Coleman1967}, which considers a single continuum body $\Omega$ subjected external source of force, heat flux, and entropy flux, we follow the method proposed by Bouchbinder \cite{Bouchbinder2009} to derive the expression of the first and second laws of thermodynamics by taking into account the effects of the reservoir, which is thermally connected with $\Omega$. Then, the first and second laws read
\begin{align}
\int_{\Omega}\dot{u}\dV +\dot{U}_{\mathrm{R}} &= \int_{\Omega} \SS:\frac{1}{2}\dot{\CC} \dV \label{eq:TD_eq_energy} \\
\int_{\Omega}\dot{s}\dV + \dot{S}_{\mathrm{R}} &\geq 0,
\label{eq:TD_eq_entropy}
\end{align}
where $u$ and $s$ denote the energy density and entropy density in $\Omega$ while $U_{\mathrm{R}}$ and ${S}_{\mathrm{R}}$ are the total energy and entropy in the reservoir, respectively. The variables $\SS$ and $\CC=\FFT\FF$ denote the Piola-Kirchhoff stress tensor and the right Cauchy-Green tensor, respectively. As the reservoir is in equilibrium state, its temperature can be defined as $T_{\mathrm{R}}=\partial U_{\mathrm{R}} / \partial {S}_{\mathrm{R}}$, resulting in the relation $\dot{S}_{\mathrm{R}}=\dot{U}_{\mathrm{R}}/T_{\mathrm{R}}$. Inserting this relation and the first law \eqref{eq:TD_eq_energy} into the second law \eqref{eq:TD_eq_entropy}, it can be transformed as 
\begin{align}
-\Bkt{1-\frac{T}{T_{\mathrm{R}}}} \dot{U}_{\mathrm{R}}
+
\int_{\Omega}\Bkt{\SS:\frac{1}{2}\dot{\CC} - \dot{\Psi} -\dot{T}s}\dV
\geq 0,
\label{eq:TD_eq_TD2}
\end{align}
where the free energy density $\Psi$ is operationally defined as $\Psi=u-Ts$ to avoid the explicit dependence of the energy on entropy $u=u(s)$.

The variable $-\dot{U}_{\mathrm{R}}$ in Equation \eqref{eq:TD_eq_TD2} equals the heat flux from the reservoir to the system $\Omega$. This process controls the temperature $T$ in $\Omega$ through adjusting $T_{\mathrm{R}}$, which is independent of what happens in $\Omega$. Therefore, the inequality \eqref{eq:TD_eq_TD2} must be satisfied for arbitrary $\dot{U}_{\mathrm{R}}$, requiring 
\begin{align}
-\Bkt{1-\frac{T}{T_{\mathrm{R}}}} \dot{U}_{\mathrm{R}}
\geq 0,
\label{eq:TD_eq_TD2_UR}
\end{align}
which can be fulfilled by specifying the direction of the heat flux $Q \coloneqq -\dot{U}_{\mathrm{R}}= - K \Bkt{T-T_{\mathrm{R}}}$ with an effective scalar coefficient $K>0$. The second term in Equation \eqref{eq:TD_eq_TD2} also must be satisfied for all possible $\dot{\CC}$ and $\dot{T}$, independent of the heat heat flux into $\Omega$, and also for arbitrary $\Omega$, leading to the inequality
\begin{align}
\SS:\frac{1}{2}\dot{\CC} - \dot{\Psi} - s\dot{T} \geq 0, \label{eq:TD_ineq_Psi}
\end{align}
essentially identical to the conventional Clausius-Duhem inequality.

Based on the MD simulation results discussed in Section \ref{sec:MD} and \ref{sec:decomposition}, the free energy density $\Psi$ is assumed to be a function of temperature $T$, the right Cauchy-Green tensor $\CC$, the visco-plastic part of the deformation gradient $\FFp$ for the pair contributions, the stretch resisting part of the deformation gradient $\FFr$ for the contribution from bonded interactions, and an internal variable $\AA$ associated with the back stress, which has been mostly assumed to represent the orientation of polymer chains and is symmetric and unimodular \cite{Anand2009}. The rate of the free energy is then expressed as
\begin{align}
\dot{\Psi} 
&= \frac{\partial \Psi}{\partial T}\dot{T} + 
\frac{\partial \Psi}{\partial \CC}:\dot{\CC} +
\frac{\partial \Psi}{\partial \FFp}:\dot{\FF}^{\mathrm{p}}
+
\frac{\partial \Psi}{\partial \FFr}:\dot{\FF}^{\mathrm{r}}
+
\frac{\partial \Psi}{\partial \AA}:\dot{\AA} \nonumber\\
&=
\frac{\partial \Psi}{\partial T}\dot{T} +
2\frac{\partial \Psi}{\partial \CC}:\frac{1}{2}\dot{\CC} 
- 2\CCe\frac{\partial \Psi}{\partial \CCe}:\DDp
- 2\CCl\frac{\partial \Psi}{\partial \CCl}:\DDr
+ \frac{\partial \Psi}{\partial \AA}:\dot{\AA},
\label{eq:TD_Psi_dot}
\end{align}
where $\DDp$ is the symmetric part of the velocity gradient $\llt^{\mathrm{p}}=\dot{\FF}^{\mathrm{p}}\FF^{\mathrm{p-1}}$.
Substituting Equation \eqref{eq:TD_Psi_dot} into \eqref{eq:TD_ineq_Psi}, the Clausius-Duhem inequality becomes
\begin{align}
&\Bkt{\SS-2\frac{\partial \Psi}{\partial \CC}}:\frac{1}{2}\dot{\CC}
-\Bkt{s+\frac{\partial \Psi}{\partial T}}\dot{T}
+ 2\CCe\frac{\partial \Psi}{\partial \CCe}:\DDp 
\nonumber\\
&
+ 2\CCl\frac{\partial \Psi}{\partial \CCl}:\DDr
- \frac{\partial \Psi}{\partial \AA}:\dot{\AA}
\geq 0,
\label{eq:TD_ineq_CD_all_terms}
\end{align}
which must be satisfied for all possible $\dot{\CC}$ and $\dot{T}$, resulting in the constitutive model
\begin{align}
\SS&=2\frac{\partial \Psi}{\partial \CC}, \label{eq:CM_form_S} \\
s&=-\frac{\partial \Psi}{\partial T} \label{eq:CM_form_entropy}.
\end{align}
These expressions simplifies the inequality \eqref{eq:TD_ineq_CD_all_terms} to
\begin{align}
2\CCe\frac{\partial \Psi}{\partial \CCe}:\DDp 
+ 2\CCl\frac{\partial \Psi}{\partial \CCl}:\DDr
- \frac{\partial \Psi}{\partial \AA}:\dot{\AA}
\geq 0.
\label{eq:TD_ineq_CD_simplified}
\end{align}
As assumed that the pair and bonded contributions are independent, the inequality \eqref{eq:TD_ineq_CD_simplified} must be satisfied for all possible $\DDp$, requiring that the first term in Equation \eqref{eq:TD_ineq_CD_simplified} to be non-negative. This can be realized by defining the flow law as
\begin{align}
\DDp = \frac{1}{2\eta^{\mathrm{p}}} \underbrace{2\CCe\frac{\partial \Psi}{\partial \CCe}}_{\eqqcolon \MM^{\mathrm{pair}}} 
=
\frac{\MM^{\mathrm{pair}}}{2\eta^{\mathrm{p}}}
\label{eq:CM_form_Di}
\end{align}
with the viscosity $\eta^{\mathrm{p}}>0$, where $\MM^{\mathrm{pair}}$ denotes the pair Mandel stress tensor.

For the same reason, the last two terms in Equation \eqref{eq:TD_ineq_CD_simplified} must also be non-negative as they are variables associated with the bonded contributions. Adopting the following evolution equation for $\AA$ 
\begin{align}
\dot{\AA} = \AA\DDr+\DDr\AA-\ff(\AA)\AA
\label{eq:CM_form_dotA}
\end{align}
as widely used \cite{Anand2009,Ames2009,Xiao2019}, where the function $\ff(\AA)$ represents the recovery relaxation of $\AA$ due to the back stress originated from $\AA$. The variable ${\AA}$ would be associated with the left Cauchy-Green tensor $\bb^{\mathrm{r}}=\FFr\FF^{\mathrm{rT}}$ if the recovery term $f(\AA)$ vanishes \cite{Anand2009}. Using the equations of motion \eqref{eq:CM_form_dotA}, the last two terms in inequality \eqref{eq:TD_ineq_CD_simplified} can be simplified as
\begin{align}
\Bkt{ 2\CCl\frac{\partial \Psi}{\partial \CCl}-2\AA \frac{\partial \Psi}{\partial \AA} }:\DDr
+ \AA \frac{\partial \Psi}{\partial \AA}:\ff(\AA)
\geq 0.
\label{eq:TD_ineq_CD_simplified_bonded}
\end{align}
The simplified inequality \eqref{eq:TD_ineq_CD_simplified_bonded} can be satisfied by requiring a stronger form that each term is non-negative. Similar to Ref. \cite{Xiao2019}, we adopt the simplest form that
\begin{align}
\DDr &= \frac{1}{2\eta^{\mathrm{b}}} \underbrace{\Bkt{2\CCl\frac{\partial \Psi}{\partial \CCl}-2\AA \frac{\partial \Psi}{\partial \AA} }}_{\eqqcolon \MM^{\mathrm{eff}}} = \frac{\MM^{\mathrm{eff}}}{2\eta^{\mathrm{b}}}, 
\label{eq:CM_form_Dr}\\
\ff(\AA) &= \frac{1}{2\eta^{\mathrm{ori}}} \underbrace{2\AA \frac{\partial \Psi}{\partial \AA}}_{\eqqcolon \MM^{\mathrm{back}}} = \frac{\MM^{\mathrm{back}}}{ 2\eta^{\mathrm{ori}}}
\label{eq:CM_form_fA}
\end{align}
with the viscosity $\eta^{\mathrm{b}}>0$ and $\eta^{\mathrm{ori}}>0$, as well as the Mandel back stress $\MM^{\mathrm{back}}$ and the Mandel effective stress $\MM^{\mathrm{eff}}$. The equilibrium relations \eqref{eq:CM_form_S}, \eqref{eq:CM_form_entropy} and the evolution equations \eqref{eq:CM_form_Di},  \eqref{eq:CM_form_Dr},  \eqref{eq:CM_form_dotA}, as well as the expression for the recovery function \eqref{eq:CM_form_fA} constitute the formulation of the constitutive model satisfying the thermodynamic laws.

\subsection{Individual models}\label{subsec:CM_individual_models}
We assume the free energy density $\Psi$ can be decomposed into a thermal part $\Psi^{\mathrm{th}}(T)$, an equilibrium part $\Psie(\CC)$, an inequilibrium part accounting for the pair contribution $\Psi^{\mathrm{pair}}(\CCe)$, an inequilibrium part for the bonded contribution $\Psi^{\mathrm{bond}}(\CCl)$, and an orientation part $\Psi^{\mathrm{ori}}(\AA)$ as
\begin{align}
\Psi=\Psi^{\mathrm{th}}(T)+\Psie(\CC)+\Psi^{\mathrm{pair}}(\CCe)+\Psi^{\mathrm{bond}}(\CCl)+\Psi^{\mathrm{ori}}(\AA).
\end{align}

The expression for $\Psi^{\mathrm{th}}$ is not of interest in the present paper as we focus on the simulations under isothermal conditions. The equilibrium term is assumed to be only dependent on the volume change as there are no permanent cross-links in the present MD models for thermoplastics. The pair term $\Psi^{\mathrm{pair}}$ is assumed to be subjected to the NH model for simplicity. The bonded term is also considered as an NH model as justified in Section \ref{subsec:MD_quantitative_effects_lb_max}, specifically, as evident in Figure \ref{fig:fig_LS_MD_curve_micro_local_dev}. The expression of $\Psi^{\mathrm{ori}}$ is assumed to be a Langevin-related model as suggested by the results in Figure \ref{fig:fig_LS_MD_curve_decomposition_stress_eff_back}(b), where we take the widely used Eight-Chain model \cite{Arruda1993}. Therefore, the individual models can be summarized as
\begin{align}
\Psie &= \frac{{\kappa^{\mathrm{e}}}}{2}\Bkt{J-1}^2, \\
\Psi^{\mathrm{pair}} &= \frac{\mup}{2}\Bkt{\tr\bkt{{\CC^{\mathrm{e}*}}}-3} + \frac{{\kappa^{\mathrm{p}}}}{2}\Bkt{J^{\mathrm{e}}-1}^2, \\
\Psi^{\mathrm{bonded}} &= \frac{\mub}{2}\Bkt{\tr\bkt{\CC^{\mathrm{l}*}}-3} + \frac{{\kappa^{\mathrm{b}}}}{2}\Bkt{J^{\mathrm{b}}-1}^2, \\
\Psi^{\mathrm{ori}} &= \mu^{\mathrm{back}} \lambda_{\mathrm{L}}\Bkt{\lambda^* \beta + \lambda_{\mathrm{L}} \ln\bkt{\frac{\beta}{\sinh\,\beta}}}-\Psi^{\mathrm{ori}}_0
\end{align}
with ${\lambda^*}=\sqrt{\tr(\AA^*)/3}$, $\beta=\LLinv\bkt{{\lambda^*}/\lambda^\mathrm{L}}$ and a constant $\Psi^{\mathrm{ori}}_0$. The parameter $\lambda_{\mathrm{L}}$ represents the locking stretch while the function $\LLinv$ is the inverse of the Langevin function $\mathcal{L}(\bullet)=\coth(\bullet)-(\bullet)^{-1}$. The coefficient $\mu^{\mathrm{back}}$ can be considered as shear modulus accounting for the chain orientation. As $J^\mathrm{e}=J^{\mathrm{b}}=J$, the effective total bulk modulus can be defined as $\kappa=\kappa^{\mathrm{e}}+\kappa^{\mathrm{p}}+\kappa^{\mathrm{b}}$.

Utilizing the constitutive relation \eqref{eq:CM_form_S}, the Cauchy stress tensor can be calculated as
\begin{align}
\sigmaa = \frac{1}{J}\FF\SS\FFT =
\underbrace{\frac{\mup}{J}\dev\bkt{\bb^{\mathrm{e*}}}}_{\eqqcolon \sigmaa^{\mathrm{pair,dev}}} +
\underbrace{\frac{\mub}{J}\dev\bkt{\bb^{\mathrm{l*}}}}_{\eqqcolon \sigmaa^{\mathrm{bonded,dev}}} +
\underbrace{\kappa\Bkt{J-1}}_{\eqqcolon \sigmaa^{\mathrm{vol}}}.
\end{align}
The back stress reads
\begin{align}
\MM^{\mathrm{back}} = \mu^{\mathrm{back}} \frac{\LLinv\bkt{{\lambda^*}/\lambda^\mathrm{L}}}{3{\lambda^*}/\lambda^\mathrm{L}}\dev\bkt{\AA^*},
\end{align}
which recovers to the NH formulation at small ratio of ${\lambda^*}/\lambda^\mathrm{L}$ as $\LLinv(x)\approx 3x$ for sufficiently small positive $x$. The Mandel stresses associated with the bonded and pair terms are given by
\begin{align}
\MM^{\mathrm{bonded}} &= 2\CCl\frac{\partial \Psi}{\partial \CCl}
= \mub \dev\bkt{\CC^{\mathrm{l*}}}, \\
\MM^{\mathrm{pair}} &= 2\CCe\frac{\partial \Psi}{\partial \CCe}
= \mup \dev\bkt{\CC^{\mathrm{e*}}}.
\end{align}
The flow rates in the associated branches are defined in Equations \eqref{eq:CM_form_Di} and \eqref{eq:CM_form_Dr} with $\MM^{\mathrm{eff}}=\MM^{\mathrm{bonded}}-\MM^{\mathrm{back}}$, while the flow rate of $\AA$ is expressed as
\begin{align}
\dot{\AA} = \AA\DDr + \DDr\AA - \frac{\MM^{\mathrm{back}}}{2\eta^{\mathrm{ori}}}\AA.
\end{align}

As discussed in Section \ref{sec:decomposition}, the Eyring model is adopted for the viscosity associated with bonded term as in Equation \eqref{eq:Eyring_model}. The same model is also adopted for the pair term. They are expressed as
\begin{align}
\eta^{\mathrm{b}} 
&= \eta^{\mathrm{b}}_{\mathrm{g}0}a^{\mathrm{b}}(T) \frac{Q^{\mathrm{b}} m^{\mathrm{eff}}_{\mathrm{eq}}}{T} \Bkt{\sinh\bkt{\frac{Q^{\mathrm{b}} m^{\mathrm{eff}}_{\mathrm{eq}}}{T}}}^{-1}, \\
\eta^{\mathrm{p}} 
&= \eta^{\mathrm{p}}_{\mathrm{g}0}a^{\mathrm{p}}(T) \frac{Q^{\mathrm{p}} m^{\mathrm{pair}}_{\mathrm{eq}}}{T} \Bkt{\sinh\bkt{\frac{Q^{\mathrm{p}} m^{\mathrm{pair}}_{\mathrm{eq}}}{T}}}^{-1}
\end{align}
with $Q^{\mathrm{b}}$, $\eta^{\mathrm{b}}_{\mathrm{g}0}$, and $m^{\mathrm{eff}}_{\mathrm{eq}}$ being the specific activation volume, the stress-independent viscosity at $\Tg$, and the norm of $\MM^{\mathrm{eff}}$, respectively, associated with the bonded term. Accordingly, the quantities $Q^{\mathrm{p}}$, $\eta^{\mathrm{b}}_{\mathrm{g}0}$, and $m^{\mathrm{pair}}_{\mathrm{eq}}$ have the same meaning associated with the pair term. The Arrhenius function is taken for both terms as
\begin{align}
{a^{\mathrm{b}}(T)} = \exp\bkt{\frac{A^{\mathrm{b}}}{T}-\frac{A^{\mathrm{b}}}{\Tg}}, \\
{a^{\mathrm{p}}(T)} = \exp\bkt{\frac{A^{\mathrm{p}}}{T}-\frac{A^{\mathrm{p}}}{\Tg}}
\end{align}
with constant parameters $A^{\mathrm{b}}>0$ and $A^{\mathrm{p}}>0$. The viscosity associated with chain orientation $\eta^{\mathrm{ori}}$ is assumed to be a function of $\eta^{\mathrm{b}}$ and the shape of the system, taking the form
\begin{align}
\eta^{\mathrm{ori}} = c\, \eta^{\mathrm{b}} \Bkt{\frac{\lambda^*_{\max}}{\lambda^*_{\min}}}^m 
\end{align}
with $\lambda^*_{\max} = \max\left\{\lambda^*_1,\lambda^*_2,\lambda^*_3\right\}$ and $\lambda^*_{\min} = \min\left\{\lambda^*_1,\lambda^*_2,\lambda^*_3\right\}$, where $\lambda^{*2}_1$, $\lambda^{*2}_2$, $\lambda^{*2}_3$ denote the eigenvalues of $\CC^*$. The coefficient $c$ is a constant parameter denoting the ratio between $\eta^{\mathrm{ori}}$ and $\eta^{\mathrm{b}}$. The exponent $m$ accounts for the effect of shape on the viscosity for chain orientation, which increases with the stretch of polymer chains.

\subsection{Parameter identification}\label{subsec:CM_PI}
The parameters of the constitutive model are listed in Table \ref{tab:parameters} together with their identified values, categorized into the groups of pair, bonded, and back stresses, as well as the bulk modulus for the total stress. The bulk modulus $\kappa$ is roughly estimated according to the response of the pressure to the volume change $J$, which is not a main focuses of the present paper. The parameters in the pair group are identified according to the stress-strain curves of the pair contributions of the virial stress as presented in Section \ref{subsec:MD_primary_contributions}, where the shear modulus $\mup$ is estimated to be the slope of the stress-strain curves at small strains while the remaining parameters are identified using the same process as discussed in Section \ref{subsec:decomposition_yielding}. In the bonded group, the shear modulus $\mub$ is assumed to be identical to be the bond shear modulus fitted in Section \ref{subsec:MD_quantitative_effects_lb_max} while the activation temperature $A^{\mathrm{b}}$ and the specific activation volume $Q^{\mathrm{b}}$ are adopted from the fitted values presented in Section \ref{subsec:decomposition_yielding}. However, this process is based on the assumption that the back stress vanishes before yielding, which could result in deviations between the results from the constitutive model and the MD simulations. We attempt to reduce this deviation by adjusting as few parameters as possible. Here, only the static viscosity $\eta^{\mathrm{b}}_{\mathrm{g}0}$ is set to be adjustable. In the back group, we assume the shear modulus $\mu^{\mathrm{back}}$ is identical to $\mub$ as they originates from the same chain segments while the remaining three parameters have to be identified by fitting the stress-strain curves.

\begin{table}[htbp]
\centering
\caption{Parameters of the constitutive model and their identified values.}
\label{tab:parameters}
\begin{tabular}{l l l l l} 
\toprule
Parameter & value & unit & physical meaning & stress groups\\
\hline
$\kappa$ & 300 & $[\si{\mega\pascal}]$ & total bulk modulus & total\\
\hline
$\mup$ & 148 & $[\si{\mega\pascal}]$ & shear modulus & \multirow{4}{*}{pair}\\
$\eta^{\mathrm{p}}_{\mathrm{g}0}$ & 0.003 & $[\si{\mega\pascal\second}]$ & static viscosity at $\Tg$ & \\
$A^{\mathrm{p}}$ & 3393 & $[\si{\kelvin}]$ & activation temperature & \\
$Q^{\mathrm{p}}$ &118  & $[\si{\kelvin\per\mega\pascal}]$ & spcific activation volume & \\
\hline
$\mub$ &131 & $[\si{\mega\pascal}]$ & shear modulus & \multirow{4}{*}{bonded} \\
$\eta^{\mathrm{b}}_{\mathrm{g}0}$ & 2.686E-8 & $[\si{\mega\pascal\second}]$ & static viscosity at $\Tg$ &\\
$A^{\mathrm{b}}$ & 1008& $[\si{\kelvin}]$ & activation temperature &\\
$Q^{\mathrm{b}}$ & 80.54& $[\si{\kelvin\per\mega\pascal}]$  & spcific activation volume &\\
\hline
$\mu^{\mathrm{back}}$ &131 & $[\si{\mega\pascal}]$ & shear modulus  & \multirow{3}{*}{back}\\
$\lambda_{\mathrm{L}}$ &1.1 & $[-]$ & locking stretch &\\
$c$ & 0.049& $[-]$ & viscosity ratio &\\
$m$ & 2.39& $[-]$ &  shape exponent&\\
\bottomrule
\end{tabular}
\end{table}

\begin{figure}[h!]
\centering
\includegraphics[width=0.6\textwidth]{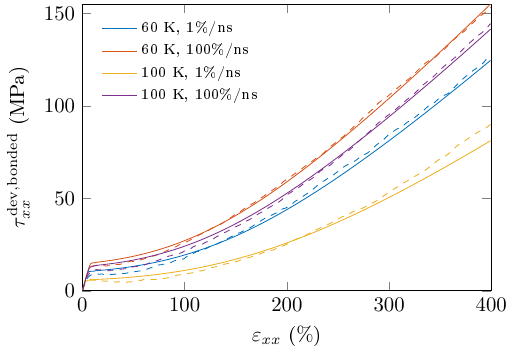}
\caption{Parameter identification: Bonded Kirchhoff stress during deformation of the constitutive model (solid curves) and MD simulations (dashed curves).}
\label{fig:fig_LS_Model_ss_PI_bond}
\end{figure}

Specifically, the parameters $\eta^{\mathrm{b}}_{\mathrm{g}0}$, $\lambda_{\mathrm{L}}$, $c$, and $m$ are identified by minimizing the deviations of the deviatoric part of the bonded stresses between the constitutive model and the MD simulation results. We consider the conditions at temperatures of $60\:\si{\kelvin}$ and $100\:\si{\kelvin}$, each for strain rates of $1\%/\mathrm{ns}$ and $1\%/\mathrm{ns}$. As shown in Figure \ref{fig:fig_LS_Model_ss_PI_bond}, the constitutive model can well reproduce the MD results at $60\:\si{\kelvin}$ and at the strain rate of $100\%/\mathrm{ns}$ for $100\:\si{\kelvin}$. For the simulation at $100\:\si{\kelvin}$ and $1\%/\mathrm{ns}$, the deviations are small at intermediate strains but start increasing from the strain of $300\%$, which might be caused by the effects of disentanglement as discussed in Section \ref{subsec:MD_primary_contributions}. As the main focus of the paper is the mechanisms of local load-bearing stretch on strain hardening, the effects of disentanglement are not further discussed here and will be systematically studied in future research.

\subsection{Validation}\label{subsec:CM_validation}
We compare the stress responses of the constitutive model with identified parameters summarized in Table \ref{tab:parameters} and the MD simulation results in uniaxial tension at temperatures of $T=20\:\si{\kelvin}-120\:\si{\kelvin}$ for every $20\:\si{\kelvin}$ below $\Tg$ with true strain rates of $\dot{\epsilon}=1\%/\mathrm{ns}-100\%/\mathrm{ns}$. It is noticeable that the temperatures considered in the validation are beyond the limits of temperatures used for parameter identification to challenge the constitutive model.

\begin{figure}[h!]
\centering
\begin{subfigure}[t]{0.48\textwidth}
\centering
\caption{$\dot{\epsilon}=1\%/\mathrm{ns}$}
\includegraphics[width=0.92\textwidth]{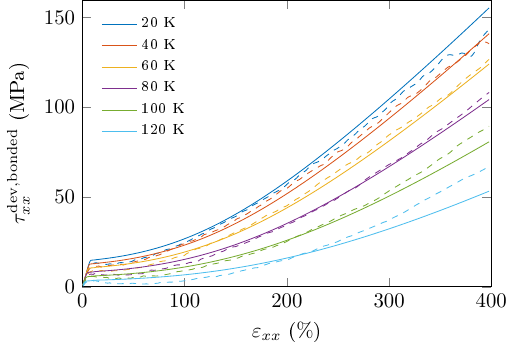}
\end{subfigure}%
\hfill
\begin{subfigure}[t]{0.48\textwidth}
\centering
\caption{$\dot{\epsilon}=10\%/\mathrm{ns}$}
\includegraphics[width=0.92\textwidth]{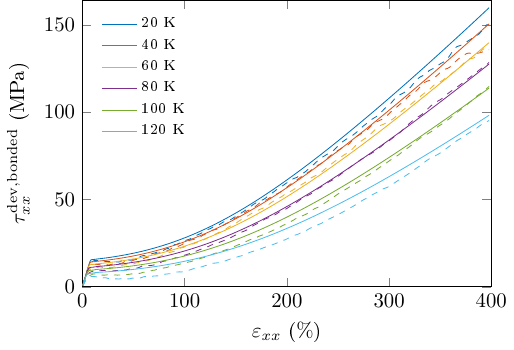}
\end{subfigure}%
\\
\begin{subfigure}[t]{0.48\textwidth}
\centering
\caption{$\dot{\epsilon}=50\%/\mathrm{ns}$}
\includegraphics[width=0.92\textwidth]{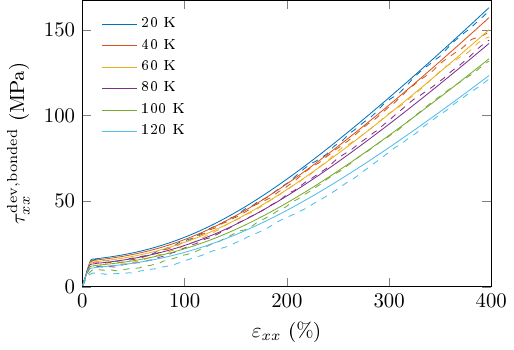}
\end{subfigure}%
\hfill
\begin{subfigure}[t]{0.48\textwidth}
\centering
\caption{$\dot{\epsilon}=100\%/\mathrm{ns}$}
\includegraphics[width=0.92\textwidth]{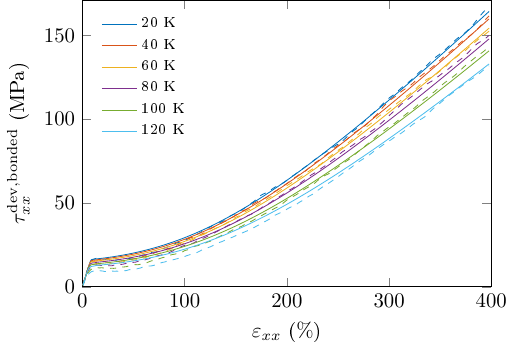}
\end{subfigure}%
\caption{Comparison between the bonded Kirchhoff stress during deformation of the constitutive model (solid curves) and MD simulations (dashed curves) at different strain rates and temperatures.}
\label{fig:fig_LS_Model_ss_validation_bond}
\end{figure}

Firstly, the bonded stresses are compared as illustrated in Figure \ref{fig:fig_LS_Model_ss_validation_bond}, demonstrating an excellent agreement between the constitutive model and the MD results, except for the deviations at the temperatures of $100\:\si{\kelvin}$ and $120\:\si{\kelvin}$ with strain rates of $1\%/\mathrm{ns}$ as shown in Figure \ref{fig:fig_LS_Model_ss_validation_bond}(a). This can be attributed to the effects of disentanglement that are not considered in this paper as discussed in Section \ref{subsec:CM_PI}, which will be pursued in future work.

\begin{figure}[h!]
\centering
\begin{subfigure}[t]{0.48\textwidth}
\centering
\caption{$\dot{\epsilon}=1\%/\mathrm{ns}$}
\includegraphics[width=0.92\textwidth]{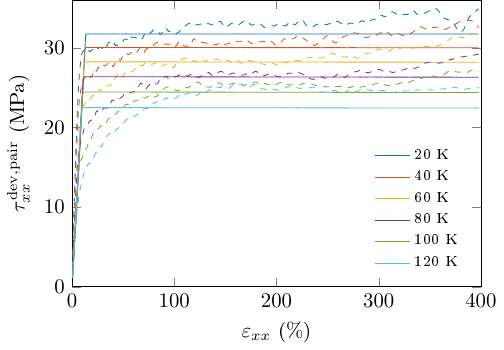}
\end{subfigure}%
\hfill
\begin{subfigure}[t]{0.48\textwidth}
\centering
\caption{$\dot{\epsilon}=1\%/\mathrm{ns}$}
\includegraphics[width=0.92\textwidth]{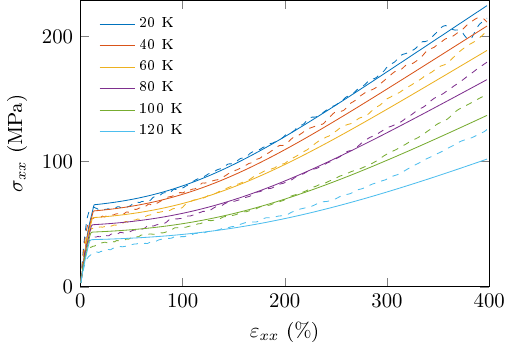}
\end{subfigure}%
\\
\begin{subfigure}[t]{0.48\textwidth}
\centering
\caption{$\dot{\epsilon}=10\%/\mathrm{ns}$}
\includegraphics[width=0.92\textwidth]{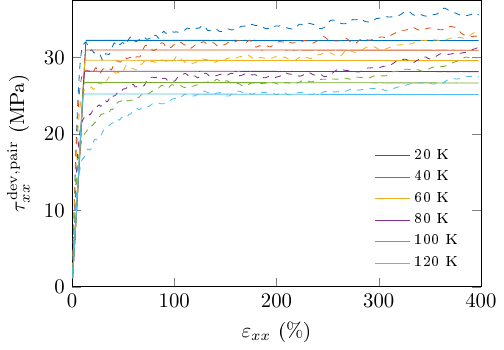}
\end{subfigure}%
\hfill
\begin{subfigure}[t]{0.48\textwidth}
\centering
\caption{$\dot{\epsilon}=10\%/\mathrm{ns}$}
\includegraphics[width=0.92\textwidth]{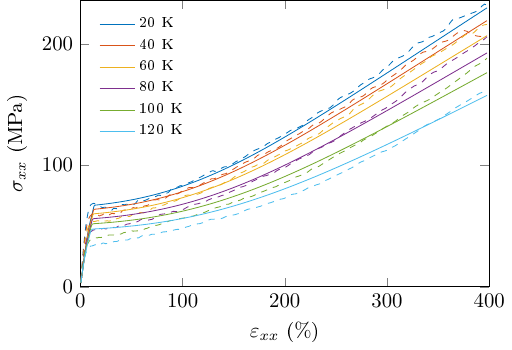}
\end{subfigure}%
\\
\begin{subfigure}[t]{0.48\textwidth}
\centering
\caption{$\dot{\epsilon}=100\%/\mathrm{ns}$}
\includegraphics[width=0.92\textwidth]{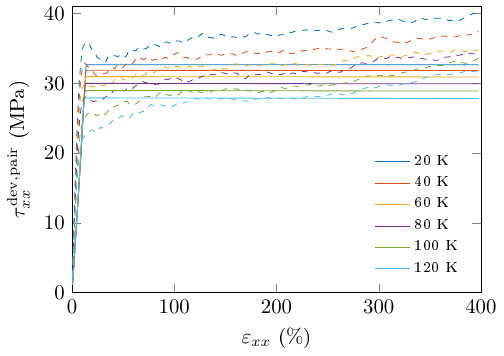}
\end{subfigure}%
\hfill
\begin{subfigure}[t]{0.48\textwidth}
\centering
\caption{$\dot{\epsilon}=100\%/\mathrm{ns}$}
\includegraphics[width=0.92\textwidth]{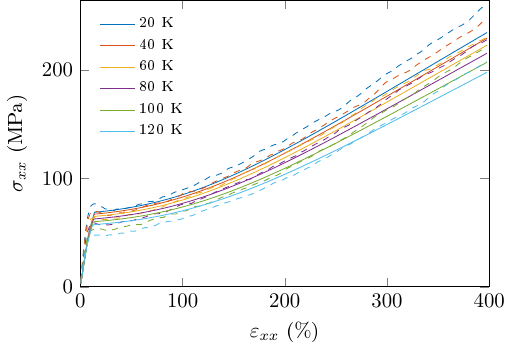}
\end{subfigure}%
\caption{Comparison between the pair Kirchhoff stress (a,c,e) and the total Cauchy stress (b,d,f) during deformation of the constitutive model (solid curves) and MD simulations (dashed curves) at different strain rates and temperatures.}
\label{fig:fig_LS_Model_ss_validation_total}
\end{figure}

Furthermore, we also present the results of the pair stress and the total stress for the completeness of evaluating the constitutive model. As shown in Figure \ref{fig:fig_LS_Model_ss_validation_total}, both the pair stress and the total stress of the constitutive model closely match the MD results. However, some deviations are noticeable for the pair stress. The deviations in the yield region could be attributed to the oversimplification of the Eyring model, which is merely a thermally-activated model and does not include structural relaxation of glassy materials. This deviation could be reduced by employing a more complex plasticity model, e.g., the phenomenological softening model \cite{Boyce1988}, the shear transformation zone (STZ) model \cite{Langer2006,Lin2023}, or the effective temperature model \cite{Kamrin2014,Xiao2015,Xiao2022}. Another deviation can be observed at large strains for simulations at high strain rates and low temperatures, where the MD results exhibit a bit strain hardening behavior. This could be interpreted by the saturation of the STZ for glassy materials, which can induce strain hardening for metallic glasses \cite{Langer2006}. However, the deviations in pair stresses is not further studied as the primary purpose of the constitutive model is to validate the mechanism of local load-bearing deformation-induced strain hardening in glassy polymers. Furthermore, the deviations resulted from pair stress at large strains are acceptable in the total stress as illustrated in Figures \ref{fig:fig_LS_Model_ss_validation_total}(b,d,f) as the bonded stress contributes significantly more than the pair stress in the hardening region.

\section{Discussion and outlook}\label{sec:conclusion}
We have conducted MD simulations to investigate the physical origin of strain hardening and its dependence on strain rate and temperature in glassy polymers. The main findings and contributions are summarized as follows:
\begin{itemize}\setlength\itemsep{0em} 
\item Primary mechanisms of strain hardening: The principal contributors to strain hardening are identified as bond and angle stresses, with negligible influence from the pair term.
\item Load-bearing bond stretches: Bond and angle stress components are governed primarily by the average stretch of the maximum bond in each polymer chain, denoted as $\langle l^{\mathrm{b}}{\max} \rangle / \langle l^{\mathrm{b}}{\max}(0) \rangle$ (referred to as load-bearing bonds), and are largely independent of strain rate and temperature in the hardening region. This suggests that elongation forces stem primarily from load-bearing bonds, while the remaining chain structure resists the stretch. 
\item Decomposition of deformation gradient: We propose decomposing the deformation gradient into a local load-bearing part, $\FFl$, and a stretch-resistance part, $\FFr$, such that $\FF = \FFl\FFr$. Here, $\FFl$ primarily contributes to stress response in the hardening region, while $\FFr$ resists chain segment elongation. 
\item Local load-bearing approximation: The local load-bearing component $\FFl$ can be approximated by $\langle l^{\mathrm{b}}{\max} \rangle / \langle l^{\mathrm{b}}{\max}(0) \rangle$ in uniaxial tension simulations. Results indicate that the bonded stress (i.e., the combined bond and angle stress) and the left Cauchy-Green tensor for $\FFl$ align well with the NH model in the hardening region, allowing us to identify the shear modulus $\mub$ from MD simulations. 
\item Constitutive modeling: Based on these mechanisms, we developed a constitutive model comprising elastic, pair, and bonded branches, which demonstrates good agreement with simulation results across a wide range of temperatures and strain rates. 
\end{itemize}

While the constitutive model presented in this paper effectively captures the temperature- and rate-dependent stress response in the hardening region at temperatures well below $\Tg$, deviations become apparent as the temperature approaches $\Tg$. This discrepancy is likely due to disentanglement effects, as indicated by the inability of polymer chain stretches to fully follow the macroscopic deformation of the entire system as discussed in Section \ref{subsec:MD_relaxation_mechanisms}. This deviation could be reduced by introducing a local deformation gradient $\FF^{\mathrm{d}}$ to account for the inelastic deformation induced by disentanglement such that $\FF=\FFl\FFr\FF^{\mathrm{d}}$, where the rate of $\FF^{\mathrm{d}}$ is driven by the bonded stress. The model incorporating the effects of disentanglement will be systematically studied in future work. Although the current constitutive model has only been validated against MD simulations, several models with a similar structure that incorporate orientation-induced back stress have been successfully validated against experimental data for a variety of glassy polymers \cite{Anand2009,Ames2009,Xiao2019}. A quantitative comparison of material parameters between real-world polymers and MD simulations will require more accurate CG MD or atomistic models.

\section*{Acknowledgment}
This work is funded by the German Research Foundation (DFG) projects 396414850 and 549959345. The author thanks Sylvain Patinet from PMMH, Paris, for insightful discussions. The author also gratefully acknowledges the support by the FAU Emerging Talents Initiative project and the HPC resources provided by the Erlangen National High Performance Computing Center (NHR@FAU) of the FAU under the NHR project b136dc. The hardware is funded by the DFG project 440719683.






%
%
%

\bibliographystyle{elsarticle-num}
\bibliography{references}

\begin{thebibliography}{10}
\expandafter\ifx\csname url\endcsname\relax
  \def\url#1{\texttt{#1}}\fi
\expandafter\ifx\csname urlprefix\endcsname\relax\def\urlprefix{URL }\fi
\expandafter\ifx\csname href\endcsname\relax
  \def\href#1#2{#2} \def\path#1{#1}\fi

\bibitem{Wendlandt2005}
M.~Wendlandt, T.~A. Tervoort, U.~W. Suter, Non-linear, rate-dependent
  strain-hardening behavior of polymer glasses, Polymer 46~(25) (2005)
  11786--11797.
\newblock \href {https://doi.org/10.1016/j.polymer.2005.08.079}
  {\path{doi:10.1016/j.polymer.2005.08.079}}.

\bibitem{Senden2010}
D.~J.~A. Senden, J.~A.~W. van Dommelen, L.~E. Govaert, Strain hardening and its
  relation to bauschinger effects in oriented polymers, Journal of Polymer
  Science Part B: Polymer Physics 48~(13) (2010) 1483--1494.
\newblock \href {https://doi.org/10.1002/polb.22056}
  {\path{doi:10.1002/polb.22056}}.

\bibitem{Senden2012}
D.~J.~A. Senden, S.~Krop, J.~A.~W. van Dommelen, L.~E. Govaert, Rate- and
  temperature-dependent strain hardening of polycarbonate, Journal of Polymer
  Science Part B: Polymer Physics 50~(24) (2012) 1680--1693.
\newblock \href {https://doi.org/10.1002/polb.23165}
  {\path{doi:10.1002/polb.23165}}.

\bibitem{Hoy2006}
R.~S. Hoy, M.~O. Robbins, Strain hardening of polymer glasses: Effect of
  entanglement density, temperature, and rate, Journal of Polymer Science Part
  B: Polymer Physics 44~(24) (2006) 3487--3500.
\newblock \href {https://doi.org/10.1002/polb.21012}
  {\path{doi:10.1002/polb.21012}}.

\bibitem{Hossain2010}
D.~Hossain, M.~Tschopp, D.~Ward, J.~Bouvard, P.~Wang, M.~Horstemeyer, Molecular
  dynamics simulations of deformation mechanisms of amorphous polyethylene,
  Polymer 51~(25) (2010) 6071--6083.
\newblock \href {https://doi.org/10.1016/j.polymer.2010.10.009}
  {\path{doi:10.1016/j.polymer.2010.10.009}}.

\bibitem{Treloar2005}
L.~R.~G. Treloar, The physics of rubber elasticity, Oxford : Clarendon Press ;
  New York : Oxford University Press, 2005.

\bibitem{Haward2012}
R.~N. Haward, The physics of glassy polymers, Springer Science \& Business
  Media, 2012.

\bibitem{vanMelick2003a}
H.~{van Melick}, L.~Govaert, H.~Meijer, On the origin of strain hardening in
  glassy polymers, Polymer 44~(8) (2003) 2493--2502.
\newblock \href {https://doi.org/10.1016/S0032-3861(03)00112-5}
  {\path{doi:10.1016/S0032-3861(03)00112-5}}.

\bibitem{Tian2018}
C.~Tian, R.~Xiao, J.~Guo, {An Experimental Study on Strain Hardening of
  Amorphous Thermosets: Effect of Temperature, Strain Rate, and Network
  Density}, Journal of Applied Mechanics 85~(10), 101012 (07 2018).
\newblock \href {https://doi.org/10.1115/1.4040692}
  {\path{doi:10.1115/1.4040692}}.

\bibitem{Kroener1959}
E.~Kr{\"o}ner, Allgemeine kontinuumstheorie der versetzungen und
  eigenspannungen, Archive for Rational Mechanics and Analysis 4~(273) (1959).
\newblock \href {https://doi.org/10.1007/BF00281393}
  {\path{doi:10.1007/BF00281393}}.

\bibitem{Lee1969}
E.~H. Lee, {Elastic-Plastic Deformation at Finite Strains}, Journal of Applied
  Mechanics 36~(1) (1969) 1--6.
\newblock \href {https://doi.org/10.1115/1.3564580}
  {\path{doi:10.1115/1.3564580}}.

\bibitem{Arruda1993}
E.~M. Arruda, M.~C. Boyce, H.~Quintus-Bosz, Effects of initial anisotropy on
  the finite strain deformation behavior of glassy polymers, International
  Journal of Plasticity 9~(7) (1993) 783--811.
\newblock \href {https://doi.org/10.1016/0749-6419(93)90052-R}
  {\path{doi:10.1016/0749-6419(93)90052-R}}.

\bibitem{Anand2003}
L.~Anand, M.~E. Gurtin, A theory of amorphous solids undergoing large
  deformations, with application to polymeric glasses, International Journal of
  Solids and Structures 40~(6) (2003) 1465--1487.
\newblock \href {https://doi.org/10.1016/S0020-7683(02)00651-0}
  {\path{doi:10.1016/S0020-7683(02)00651-0}}.

\bibitem{Anand2009}
L.~Anand, N.~M. Ames, V.~Srivastava, S.~A. Chester, A thermo-mechanically
  coupled theory for large deformations of amorphous polymers. part i:
  Formulation, International Journal of Plasticity 25~(8) (2009) 1474--1494.
\newblock \href {https://doi.org/10.1016/j.ijplas.2008.11.004}
  {\path{doi:10.1016/j.ijplas.2008.11.004}}.

\bibitem{Zhao2024a}
W.~Zhao, P.~Steinmann, S.~Pfaller, Modeling steady state rate- and
  temperature-dependent strain hardening behavior of glassy polymers, Mechanics
  of Materials (2024) 105044\href
  {https://doi.org/10.1016/j.mechmat.2024.105044}
  {\path{doi:10.1016/j.mechmat.2024.105044}}.

\bibitem{Voyiadjis2016}
G.~Z. Voyiadjis, A.~Samadi-Dooki, {Constitutive modeling of large inelastic
  deformation of amorphous polymers: Free volume and shear transformation zone
  dynamics}, Journal of Applied Physics 119~(22) (2016) 225104.
\newblock \href {https://doi.org/10.1063/1.4953355}
  {\path{doi:10.1063/1.4953355}}.

\bibitem{Zhu2022}
P.~Zhu, J.~Lin, R.~Xiao, H.~Zhou, Unravelling physical origin of the
  bauschinger effect in glassy polymers, Journal of the Mechanics and Physics
  of Solids 168 (2022) 105046.
\newblock \href {https://doi.org/10.1016/j.jmps.2022.105046}
  {\path{doi:10.1016/j.jmps.2022.105046}}.

\bibitem{Lin2022}
J.~Lin, P.~Zhu, C.~Tian, H.~Zhou, R.~Xiao, Physically-based interpretation of
  abnormal stress relaxation response in glassy polymers, Extreme Mechanics
  Letters 52 (2022) 101667.
\newblock \href {https://doi.org/10.1016/j.eml.2022.101667}
  {\path{doi:10.1016/j.eml.2022.101667}}.

\bibitem{Jatin2014}
Jatin, V.~Sudarkodi, S.~Basu, Investigations into the origins of plastic flow
  and strain hardening in amorphous glassy polymers, International Journal of
  Plasticity 56 (2014) 139--155.
\newblock \href {https://doi.org/10.1016/j.ijplas.2013.11.007}
  {\path{doi:10.1016/j.ijplas.2013.11.007}}.

\bibitem{BauwensCrowet1969}
C.~Bauwens-Crowet, J.~C. Bauwens, G.~Hom{\`{e}}s, Tensile yield-stress behavior
  of glassy polymers, Journal of Polymer Science Part A-2: Polymer Physics
  7~(4) (1969) 735--742.
\newblock \href {https://doi.org/10.1002/pol.1969.160070411}
  {\path{doi:10.1002/pol.1969.160070411}}.

\bibitem{BauwensCrowet1973}
C.~Bauwens-Crowet, The compression yield behaviour of polymethyl methacrylate
  over a wide range of temperatures and strain-rates, Journal of Materials
  Science 8~(7) (1973) 968--979.
\newblock \href {https://doi.org/10.1007/BF00756628}
  {\path{doi:10.1007/BF00756628}}.

\bibitem{Siviour2005}
C.~Siviour, S.~Walley, W.~Proud, J.~Field, The high strain rate compressive
  behaviour of polycarbonate and polyvinylidene difluoride, Polymer 46~(26)
  (2005) 12546--12555.
\newblock \href {https://doi.org/10.1016/j.polymer.2005.10.109}
  {\path{doi:10.1016/j.polymer.2005.10.109}}.

\bibitem{Diani2015}
J.~Diani, P.~Gilormini, J.~Arrieta, Direct experimental evidence of
  time-temperature superposition at finite strain for an amorphous polymer
  network, Polymer 58 (2015) 107--112.
\newblock \href {https://doi.org/10.1016/j.polymer.2014.12.045}
  {\path{doi:10.1016/j.polymer.2014.12.045}}.

\bibitem{Federico2018}
C.~Federico, J.~Bouvard, C.~Combeaud, N.~Billon, Large strain/time dependent
  mechanical behaviour of pmmas of different chain architectures. application
  of time-temperature superposition principle, Polymer 139 (2018) 177--187.
\newblock \href {https://doi.org/10.1016/j.polymer.2018.02.021}
  {\path{doi:10.1016/j.polymer.2018.02.021}}.

\bibitem{Bernard2020}
C.~Bernard, N.~Bahlouli, D.~George, Y.~R{\'{e}}mond, S.~Ahzi, Identification of
  the dynamic behavior of epoxy material at large strain over a wide range of
  temperatures, Mechanics of Materials 143 (2020) 103323.
\newblock \href {https://doi.org/10.1016/j.mechmat.2020.103323}
  {\path{doi:10.1016/j.mechmat.2020.103323}}.

\bibitem{Xiao2019}
R.~Xiao, C.~Tian, A constitutive model for strain hardening behavior of
  predeformed amorphous polymers: Incorporating dissipative dynamics of
  molecular orientation, Journal of the Mechanics and Physics of Solids 125
  (2019) 472--487.
\newblock \href {https://doi.org/10.1016/j.jmps.2019.01.008}
  {\path{doi:10.1016/j.jmps.2019.01.008}}.

\bibitem{Qian2008}
H.-J. Qian, P.~Carbone, X.~Chen, H.~A. Karimi-Varzaneh, C.~C. Liew,
  F.~M{\"{u}}ller-Plathe, Temperature-transferable coarse-grained potentials
  for ethylbenzene, polystyrene, and their mixtures, Macromolecules 41~(24)
  (2008) 9919--9929.
\newblock \href {https://doi.org/10.1021/ma801910r}
  {\path{doi:10.1021/ma801910r}}.

\bibitem{Reith2003}
D.~Reith, M.~P{\"{u}}tz, F.~M{\"{u}}ller-Plathe, Deriving effective mesoscale
  potentials from atomistic simulations, Journal of Computational Chemistry
  24~(13) (2003) 1624--1636.
\newblock \href {https://doi.org/10.1002/jcc.10307}
  {\path{doi:10.1002/jcc.10307}}.

\bibitem{Depa2005}
P.~K. Depa, J.~K. Maranas, {Speed up of dynamic observables in coarse-grained
  molecular-dynamics simulations of unentangled polymers}, The Journal of
  Chemical Physics 123~(9) (2005) 094901.
\newblock \href {https://doi.org/10.1063/1.1997150}
  {\path{doi:10.1063/1.1997150}}.

\bibitem{Rahimi2012}
M.~Rahimi, I.~Iriarte-Carretero, A.~Ghanbari, M.~C. B{\"{o}}hm,
  F.~M{\"{u}}ller-Plathe, Mechanical behavior and interphase structure in a
  silica{\textendash}polystyrene nanocomposite under uniaxial deformation,
  Nanotechnology 23~(30) (2012) 305702.
\newblock \href {https://doi.org/10.1088/0957-4484/23/30/305702}
  {\path{doi:10.1088/0957-4484/23/30/305702}}.

\bibitem{Lyulin2002}
A.~V. Lyulin, M.~Michels, Molecular dynamics simulation of bulk atactic
  polystyrene in the vicinity of tg, Macromolecules 35~(4) (2002) 1463--1472.
\newblock \href {https://doi.org/10.1021/ma011318u}
  {\path{doi:10.1021/ma011318u}}.

\bibitem{Kaliappan2005}
S.~K. Kaliappan, B.~Cappella, Temperature dependent elastic--plastic behaviour
  of polystyrene studied using afm force--distance curves, Polymer 46~(25)
  (2005) 11416--11423.
\newblock \href {https://doi.org/10.1016/j.polymer.2005.09.066}
  {\path{doi:10.1016/j.polymer.2005.09.066}}.

\bibitem{Xia2017}
W.~Xia, J.~Song, C.~Jeong, D.~D. Hsu, F.~R.~J. Phelan, J.~F. Douglas, S.~Keten,
  Energy-renormalization for achieving temperature transferable coarse-graining
  of polymer dynamics, Macromolecules 50~(21) (2017) 8787--8796.
\newblock \href {https://doi.org/10.1021/acs.macromol.7b01717}
  {\path{doi:10.1021/acs.macromol.7b01717}}.

\bibitem{Xia2019}
W.~Xia, N.~K. Hansoge, W.-S. Xu, F.~R. Phelan, S.~Keten, J.~F. Douglas, Energy
  renormalization for coarse-graining polymers having different segmental
  structures, Science Advances 5~(4) (2019) eaav4683.
\newblock \href {https://doi.org/10.1126/sciadv.aav4683}
  {\path{doi:10.1126/sciadv.aav4683}}.

\bibitem{Kremer1990}
K.~Kremer, G.~S. Grest, Dynamics of entangled linear polymer melts: A
  molecular-dynamics simulation, The Journal of Chemical Physics 92~(8) (1990)
  5057--5086.
\newblock \href {https://doi.org/10.1063/1.458541}
  {\path{doi:10.1063/1.458541}}.

\bibitem{Morse1929}
P.~M. Morse, Diatomic molecules according to the wave mechanics. ii.
  vibrational levels, Phys. Rev. 34 (1929) 57--64.
\newblock \href {https://doi.org/10.1103/PhysRev.34.57}
  {\path{doi:10.1103/PhysRev.34.57}}.

\bibitem{Ries2019}
M.~Ries, G.~Possart, P.~Steinmann, S.~Pfaller, Extensive {CGMD} simulations of
  atactic ps providing pseudo experimental data to calibrate nonlinear
  inelastic continuum mechanical constitutive laws, Polymers 11~(11) (2019).
\newblock \href {https://doi.org/10.3390/polym11111824}
  {\path{doi:10.3390/polym11111824}}.

\bibitem{Zhao2021}
W.~Zhao, M.~Ries, P.~Steinmann, S.~Pfaller, A viscoelastic-viscoplastic
  constitutive model for glassy polymers informed by molecular dynamics
  simulations, International Journal of Solids and Structures 226-227 (2021)
  111071.
\newblock \href {https://doi.org/10.1016/j.ijsolstr.2021.111071}
  {\path{doi:10.1016/j.ijsolstr.2021.111071}}.

\bibitem{Zhao2024}
W.~Zhao, R.~Xiao, P.~Steinmann, S.~Pfaller, Time--temperature correlations of
  amorphous thermoplastics at large strains based on molecular dynamics
  simulations, Mechanics of Materials 190 (2024) 104926.
\newblock \href {https://doi.org/10.1016/j.mechmat.2024.104926}
  {\path{doi:10.1016/j.mechmat.2024.104926}}.

\bibitem{Zhao2024b}
W.~Zhao, Y.~Jain, F.~M{\"{u}}ller-Plathe, P.~Steinmann, S.~Pfaller,
  Investigating fracture mechanisms in glassy polymers using coupled
  particle-continuum simulations, Journal of the Mechanics and Physics of
  Solids 193 (2024) 105884.
\newblock \href {https://doi.org/10.1016/j.jmps.2024.105884}
  {\path{doi:10.1016/j.jmps.2024.105884}}.

\bibitem{Ghanbari2011}
A.~Ghanbari, M.~C. B{\"{o}}hm, F.~M{\"{u}}ller-Plathe, A simple reverse mapping
  procedure for coarse-grained polymer models with rigid side groups,
  Macromolecules 44~(13) (2011) 5520--5526.
\newblock \href {https://doi.org/10.1021/ma2005958}
  {\path{doi:10.1021/ma2005958}}.

\bibitem{Zhao2023}
W.~Zhao, Multiscale modeling of the fracture behavior of glassy polymers across
  the atomistic and continuum scale, Ph.D. thesis,
  Friedrich-Alexander-Universit{\"{a}}t Erlangen-N{\"{u}}rnberg (2023).
\newblock \href {https://doi.org/10.25593/open-fau-401}
  {\path{doi:10.25593/open-fau-401}}.

\bibitem{vanMelick2003}
H.~{van Melick}, L.~Govaert, B.~Raas, W.~Nauta, H.~Meijer, Kinetics of ageing
  and re-embrittlement of mechanically rejuvenated polystyrene, Polymer 44~(4)
  (2003) 1171--1179.
\newblock \href {https://doi.org/10.1016/S0032-3861(02)00863-7}
  {\path{doi:10.1016/S0032-3861(02)00863-7}}.

\bibitem{Thompson2022}
A.~P. Thompson, H.~M. Aktulga, R.~Berger, D.~S. Bolintineanu, W.~M. Brown,
  P.~S. Crozier, P.~J. {in 't Veld}, A.~Kohlmeyer, S.~G. Moore, T.~D. Nguyen,
  R.~Shan, M.~J. Stevens, J.~Tranchida, C.~Trott, S.~J. Plimpton, Lammps - a
  flexible simulation tool for particle-based materials modeling at the atomic,
  meso, and continuum scales, Computer Physics Communications 271 (2022)
  108171.
\newblock \href {https://doi.org/10.1016/j.cpc.2021.108171}
  {\path{doi:10.1016/j.cpc.2021.108171}}.

\bibitem{Parrinello1981}
M.~Parrinello, A.~Rahman, Polymorphic transitions in single crystals: A new
  molecular dynamics method, Journal of Applied Physics 52~(12) (1981)
  7182--7190.
\newblock \href {https://doi.org/10.1063/1.328693}
  {\path{doi:10.1063/1.328693}}.

\bibitem{Martyna1994}
G.~J. Martyna, D.~J. Tobias, M.~L. Klein, Constant pressure molecular dynamics
  algorithms, The Journal of Chemical Physics 101~(5) (1994) 4177--4189.
\newblock \href {https://doi.org/10.1063/1.467468}
  {\path{doi:10.1063/1.467468}}.

\bibitem{Shinoda2004}
W.~Shinoda, M.~Shiga, M.~Mikami, Rapid estimation of elastic constants by
  molecular dynamics simulation under constant stress, Physical Review B 69
  (2004) 134103.
\newblock \href {https://doi.org/10.1103/PhysRevB.69.134103}
  {\path{doi:10.1103/PhysRevB.69.134103}}.

\bibitem{Thompson2009}
A.~P. Thompson, S.~J. Plimpton, W.~Mattson, General formulation of pressure and
  stress tensor for arbitrary many-body interaction potentials under periodic
  boundary conditions, The Journal of Chemical Physics 131~(15) (2009) 154107.
\newblock \href {https://doi.org/10.1063/1.3245303}
  {\path{doi:10.1063/1.3245303}}.

\bibitem{Zimmerman2004}
J.~A. Zimmerman, E.~B. WebbIII, J.~J. Hoyt, R.~E. Jones, P.~A. Klein, D.~J.
  Bammann, Calculation of stress in atomistic simulation, Modelling and
  Simulation in Materials Science and Engineering 12~(4) (2004) S319.
\newblock \href {https://doi.org/10.1088/0965-0393/12/4/S03}
  {\path{doi:10.1088/0965-0393/12/4/S03}}.

\bibitem{Subramaniyan2008}
A.~K. Subramaniyan, C.~Sun, Continuum interpretation of virial stress in
  molecular simulations, International Journal of Solids and Structures 45~(14)
  (2008) 4340--4346.
\newblock \href {https://doi.org/10.1016/j.ijsolstr.2008.03.016}
  {\path{doi:10.1016/j.ijsolstr.2008.03.016}}.

\bibitem{Hoy2007}
R.~S. Hoy, M.~O. Robbins, Strain hardening in polymer glasses: Limitations of
  network models, Phys. Rev. Lett. 99 (2007) 117801.
\newblock \href {https://doi.org/10.1103/PhysRevLett.99.117801}
  {\path{doi:10.1103/PhysRevLett.99.117801}}.

\bibitem{Coleman1963}
B.~D. Coleman, W.~Noll, The thermodynamics of elastic materials with heat
  conduction and viscosity, Archive for Rational Mechanics and Analysis 13~(1)
  (1963) 167--178.
\newblock \href {https://doi.org/10.1007/BF01262690}
  {\path{doi:10.1007/BF01262690}}.

\bibitem{Coleman1967}
B.~D. Coleman, M.~E. Gurtin, {Thermodynamics with Internal State Variables},
  The Journal of Chemical Physics 47~(2) (1967) 597--613.
\newblock \href {https://doi.org/10.1063/1.1711937}
  {\path{doi:10.1063/1.1711937}}.

\bibitem{Bouchbinder2009}
E.~Bouchbinder, J.~S. Langer, Nonequilibrium thermodynamics of driven amorphous
  materials. i. internal degrees of freedom and volume deformation, Phys. Rev.
  E 80 (2009) 031131.
\newblock \href {https://doi.org/10.1103/PhysRevE.80.031131}
  {\path{doi:10.1103/PhysRevE.80.031131}}.

\bibitem{Ames2009}
N.~M. Ames, V.~Srivastava, S.~A. Chester, L.~Anand, A thermo-mechanically
  coupled theory for large deformations of amorphous polymers. part ii:
  Applications, International Journal of Plasticity 25~(8) (2009) 1495--1539.
\newblock \href {https://doi.org/10.1016/j.ijplas.2008.11.005}
  {\path{doi:10.1016/j.ijplas.2008.11.005}}.

\bibitem{Boyce1988}
M.~C. Boyce, D.~M. Parks, A.~S. Argon, Large inelastic deformation of glassy
  polymers. part i: rate dependent constitutive model, Mechanics of Materials
  7~(1) (1988) 15--33.
\newblock \href {https://doi.org/10.1016/0167-6636(88)90003-8}
  {\path{doi:10.1016/0167-6636(88)90003-8}}.

\bibitem{Langer2006}
J.~Langer, Shear-transformation-zone theory of deformation in metallic glasses,
  Scripta Materialia 54~(3) (2006) 375--379, viewpoint set no: 37. On
  mechanical behavior of metallic glasses.
\newblock \href {https://doi.org/10.1016/j.scriptamat.2005.10.005}
  {\path{doi:10.1016/j.scriptamat.2005.10.005}}.

\bibitem{Lin2023}
J.~Lin, J.~Qian, Y.~Xie, J.~Wang, R.~Xiao, A mean-field shear transformation
  zone theory for amorphous polymers, International Journal of Plasticity 163
  (2023) 103556.
\newblock \href {https://doi.org/10.1016/j.ijplas.2023.103556}
  {\path{doi:10.1016/j.ijplas.2023.103556}}.

\bibitem{Kamrin2014}
K.~Kamrin, E.~Bouchbinder, Two-temperature continuum thermomechanics of
  deforming amorphous solids, Journal of the Mechanics and Physics of Solids 73
  (2014) 269--288.
\newblock \href {https://doi.org/10.1016/j.jmps.2014.09.009}
  {\path{doi:10.1016/j.jmps.2014.09.009}}.

\bibitem{Xiao2015}
R.~Xiao, T.~D. Nguyen, An effective temperature theory for the nonequilibrium
  behavior of amorphous polymers, Journal of the Mechanics and Physics of
  Solids 82 (2015) 62--81.
\newblock \href {https://doi.org/10.1016/j.jmps.2015.05.021}
  {\path{doi:10.1016/j.jmps.2015.05.021}}.

\bibitem{Xiao2022}
R.~Xiao, C.~Tian, Y.~Xu, P.~Steinmann, Thermomechanical coupling in glassy
  polymers: An effective temperature theory, International Journal of
  Plasticity 156 (2022) 103361.
\newblock \href {https://doi.org/10.1016/j.ijplas.2022.103361}
  {\path{doi:10.1016/j.ijplas.2022.103361}}.

\end{thebibliography}

\end{document}